\documentclass{JHEP3}
\usepackage{amsmath}
\usepackage{amssymb}
\usepackage{cite}

\usepackage{epsfig}

\psfull

\title{The $\mathbf{a_T}$ distribution
        of the $\mathbf{Z}$ boson at hadron colliders.}

\author{ Andrea Banfi \\
  Universit\`a degli studi di Milano-Bicocca and INFN, Sezione di Milano-Bicocca, Italy.\\
  Institute for Theoretical Physics, ETH Zurich, 8093 Zurich, Switzerland.\\
  E-mail: \email{banfi@itp.phys.ethz.ch}}

 \author{ Mrinal Dasgupta, Rosa Mar\'ia Dur\'an Delgado\\
 School of Physics and Astronomy, University of Manchester \\
 Manchester M13 9PL, U.K.\\
 E-mail: 
 \email{Mrinal.Dasgupta@manchester.ac.uk, Rosa.Duran-delgado@postgrad.manchester.ac.uk}
}

\abstract{We provide the first theoretical study of a novel variable, $a_T$,
  proposed in ref.~\cite{wyattvester} as a more accurate probe of the
  region of low transverse momentum $p_T$, for the $Z$ boson $p_T$
  distribution at hadron colliders. The $a_T$ is the component of
  $p_T$ transverse to a suitably defined axis. Our study involves
  resummation of large logarithms in $a_T$ up to the next-to--leading
  logarithmic accuracy and we compare the results to those for the
  well-known $p_T$ distribution, identifying important physical
  differences between the two cases. We also test our resummed result
  at the two-loop level by comparing its expansion to order
  $\alpha_s^2$ with the corresponding fixed-order results and find
  agreement with our expectations.}

\preprint{MAN/HEP/2009/34}

\keywords{Standard Model, NLO Computations, Hadronic Colliders, QCD}

\begin{document}

\section{Introduction}
The study of $W$ and $Z$ boson production at hadron colliders via the
Drell-Yan process~\cite{ESW} has formed a very significant part of
particle phenomenology over a period of several
years~\cite{WS,tev1,tev2}.  In the era of the LHC these studies
continue to occupy an important role for a variety of reasons. For
instance an accurate understanding of the production rates and
transverse momentum ($p_T$) distributions of the $W$ and $Z$, or
equivalently those of lepton pairs obtained from gauge boson decays,
can be used for diverse purposes.  These range from more prosaic
applications such as luminosity monitoring at the LHC to measurement
of the $W$ mass and perhaps most interestingly for discovery of new
physics, which may manifest itself via the decay of new gauge bosons
to lepton pairs.

In this paper we shall concentrate on the $p_T$ spectrum of the $Z$
boson and related quantities. The $Z$ boson $p_T$ distribution has in
fact received considerable theoretical and experimental attention in
the past but there remain aspects where it is desirable to have an
improved understanding of certain physical issues. One such important
issue is the role of the non-perturbative or ``intrinsic'' $k_T$
component which has a sizable effect on the $p_T$ spectrum at low
$p_T$ (see e.g.\ refs.~\cite{DSW,BLNY, Bozzipt}).  The fact that the
incoming quarks/anti-quarks which fuse to form gauge bosons are part
of extended objects (protons or anti-protons) and have interactions
with other constituents thereof generates a small transverse
momentum (that can be viewed as the Fermi motion of partons inside the
proton) and which a priori one might expect to be of order of the QCD
scale $\Lambda_{\mathrm{QCD}}$.

Since the intrinsic $k_T$ has a non-perturbative origin it cannot be
computed within conventional methods of perturbative QCD. One can
however model the intrinsic $k_T$ as an essentially Gaussian smearing
of perturbatively calculated $p_T$ spectra and hope to constrain the
parameters of the Gaussian by fitting the theoretical prediction to
experimental data. An example of this procedure is provided by the
work of Brock, Landry, Nadolsky and Yuan (BLNY) who proposed a
non-perturbative Gaussian form factor that in conjunction with
perturbative calculations was able to describe both Tevatron Run-1 $Z$
data as well as lower $Q^2$ Drell-Yan data~\cite{BLNY}. Alternatively
one may use a Monte Carlo event generator such as
HERWIG++~\cite{herwig} to phenomenologically investigate the same
issue.  As discussed in ref.~\cite{seygies} these studies yield
somewhat larger than expected values for the mean intrinsic $k_T$ per
parton and also reveal a dependence of this quantity on the collider
energy which features are desirable to understand better.

Additionally, as pointed out by Berge et al.~\cite{slave}, a
phenomenological study in semi-inclusive DIS processes (SIDIS) for
small Bjorken-$x$, $x < 10^{-3}$, suggests an effective dependence of
the non-perturbative form factor on $x$.\footnote{This $x$ dependence
  may be merely an effective parametrisation of missing
  \emph{perturbative} BFKL effects. Another observable, the vector
  $p_T$ of the current hemisphere in the DIS Breit frame which can be
  used to investigate this $x$ dependence, was suggested in
  ref.~\cite{DasDel06}.} Extrapolating the effect to the LHC where
such small-$x$ values become relevant, one may expect to see
significantly broader Higgs and vector boson $p_T$ spectra than one
would in the absence of small-$x$ effects.  Tevatron studies with high
rapidity vector boson samples may help to provide further information
on the role, if any, of small-$x$ broadening.

Given the importance of the studies we have mentioned above, in the
context of the LHC and the precise determination of $p_T$ spectra
there, it is important to have as thorough a probe of the low $p_T$
region of $Z$ boson production as is possible. Investigations carried
out using the conventional $Z$ boson $p_T$ spectrum mainly suffer from large
uncertainties arising from experimental systematics, dominated by
resolution unfolding and the dependence on $p_T$ of event selection
efficiencies, as discussed in detail in ref.~\cite{wyattvester}.

An interesting observable that is less sensitive to the above effects
and also essentially insensitive to the momentum resolution of leptons
produced by the decaying vector boson was proposed in ref.~\cite{wyattvester}.
This variable is just the transverse component, $a_T$, of the $p_T$
with respect to the lepton thrust axis, which we shall define more
precisely later. It has been suggested that the $a_T$ variable will be
experimentally better determined at low $a_T$ than the standard $p_T$
variable and hence it would make a more accurate probe for issues such
as our understanding of initial state radiation and the precise role
of intrinsic $k_T$, including issues such as potential small-$x$
broadening.

Before one can access information on non-perturbative effects however,
it is of vital importance to have a sound perturbative estimate of the
observable at hand. In the low $p_T$ region of interest to us one is
dealing with the emission of soft and/or collinear gluons which is
logarithmically enhanced.  The resummation of large logarithms of the
form $1/p_T \left (\alpha_s^n \ln^m M^2/p_T^2 \right )$, where $M$ is
the lepton-pair invariant mass and $m \leq 2n-1$, has been a subject
of interest over decades~\cite{DDT,APP,CSS,ERV,EV,RESBOS} and has now
been carried out to next-to-next-to leading logarithmic (NNLL)
accuracy~\cite{Catani}.  After matching such resummations with
fixed-order estimates to NLO accuracy one has a state-of--the art
theoretical prediction for the perturbative region.

In the present paper we address the issue of resummation for the $a_T$
variable. We point out that there are some similarities to resummation
for the $p_T$ distribution but also some important differences that
manifest themselves in the shape of the resummed distribution. We aim
to provide a next-to--leading logarithmic (NLL) resummation that we
envisage could be extended to NNLL level subsequently. The NLL 
resummed form we provide here can however already be used after
matching to full next-to--leading order (NLO) results for accurate 
phenomenological studies of $a_T$.

This paper is organised along the following lines: we begin by
discussing the definition of the $a_T$ and its dependence on multiple
soft gluon emission which is important at low $a_T$, where one
encounters large logarithms.  In the following section we sketch a
leading-order calculation for the $a_T$ distribution, consigning
details to an appendix, which helps to illustrate some features of the
$a_T$ distribution such as the precise origin of logarithmically
enhanced terms. In the subsequent section we carry out a resummation
of the logarithms of $a_T$ to NLL accuracy pointing out the relation
to our recent work on azimuthal jet decorrelations~\cite{BanDasDel}.
Next we identify a relationship between the $a_T$ and $p_T$
distributions at fixed order and check this relationship with the help
of a numerical fixed-order calculation using the program
MCFM~\cite{CampEll}, which is a non-trivial test of our
resummation. We conclude by pointing out the possibilities for further
work which involve a possible extension of our resummation to NNLL
accuracy (as has been done for the $p_T$ distribution~\cite{Catani})
as well as matching to the MCFM results and phenomenological
investigation once final experimental data becomes available.

\section{Definition of $\mathbf{a_T}$  and soft limit kinematics}
\label{sec:definition}
We will be concerned in this paper with large logarithms in the
perturbative description of the $a_T$ variable and their resummation.
Since these logarithms have their origin in multiple soft and/or
collinear emissions from the incoming hard partons we need to derive
the dependence of the $a_T$ on such emissions. In this section
therefore we define precisely the $a_T$ and obtain its dependence on
the small transverse momenta $k_t$ of emissions.

We recall that we are considering the production of $Z$ bosons via the
Drell-Yan (and QCD Compton) mechanisms which subsequently decay to a
lepton pair. The $a_T$ is the component of the lepton pair (or
equivalently $Z$ boson) $p_T$ transverse to a suitably defined axis.
The precise definition of the lepton thrust axis as employed in
ref.~\cite{wyattvester} is provided below:
\begin{equation}
\label{eq:axis}
\hat{n} = \frac{\vec{p}_{t1}-\vec{p}_{t2}}{|\vec{p}_{t1}-\vec{p}_{t2}|},
\end{equation}
where $\vec{p}_{t1}$ and $\vec{p}_{t2}$ are the transverse momenta of
the two leptons and thus $\hat{n}$ is a unit vector in the plane
transverse to the beam direction. It is straightforward to verify that
this is the axis with respect to which the two leptons have equal
transverse momenta.

We now consider multiple emissions from the incoming partons which
(neglecting the intrinsic $k_T$) are back-to--back along the beam
direction.  From conservation of transverse momentum we thus have
$\vec p_{t1}+\vec p_{t2}=-\sum_i \vec k_{ti}$ which means that the
lepton pair or $Z$ boson $p_T$ is just minus the vector sum of emitted
gluon transverse momenta $\vec k_{ti}$, where we refer to the momentum
transverse to the beam axis.  To obtain the dependence of $a_T$ on the
$k_{ti}$ we wish to find the component of this sum normal to the axis
defined in eq.~\eqref{eq:axis}. The axis is given by (writing $\vec
p_{t2}$ in terms of $\vec p_{t1}$ and $\vec k_{ti})$
\begin{equation}
\hat{n} = \frac{2\vec p_{t1}+\sum_i\vec k_{ti}}{|2\vec p_{t1}+\sum_i \vec k_{ti}|} 
\approx \frac{\vec p_{t1}}{|\vec p_{t1}|},
\end{equation}
where to obtain the last equation we have neglected the dependence of
the axis on emissions $k_{ti}$.  The reason for doing so is that we
are projecting the vector sum of the $k_{ti}$ along and normal to the
axis and any term $\mathcal{O}\left (k_{ti} \right)$ in the definition
of the axis impacts the projected quantity only at the level of terms
bilinear or quadratic in the small $k_{ti}$.  Such terms can be
ignored compared to the leading linear terms $\sim k_{ti}$ that we
shall retain and thus to our accuracy the axis is along the lepton
direction.\footnote{To be more precise the recoil of the axis against
  soft emissions, if retained, corrects our result only by terms that
  vanish as $a_T \to 0$. Such terms are beyond the scope of NLL
  resummation but will be included up to NLO due to the matching.}

\begin{figure}[htbp]
  \centering
  \epsfig{file=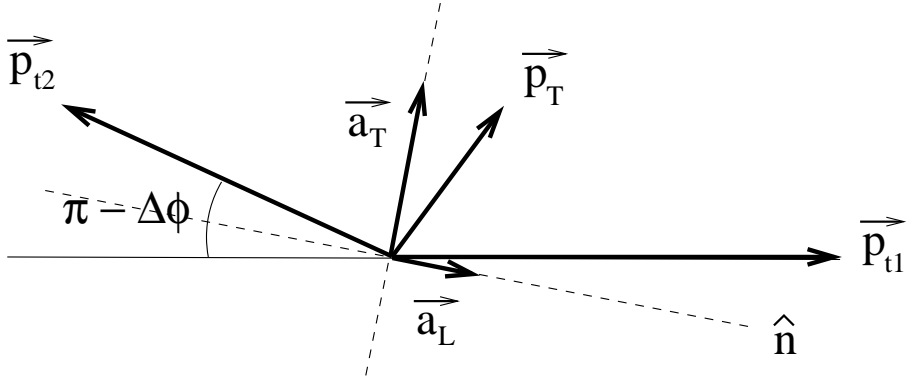,width=.8\textwidth}
  \caption{A di-lepton event in the plane transverse to the
    beam. Besides final state momenta, the two components of vector
    boson transverse momentum $\vec a_T$ and $\vec a_T$ are displayed,
    as well as $\pi-\Delta\phi$, where $\Delta\phi$ is the angle
    between the transverse momenta of the two leptons. See text for a
    full description.}
  \label{fig:at-axis}
\end{figure}

We can parametrise the lepton and gluon momenta in the plane
transverse to the beam as below:
\begin{equation}
  \begin{split}
    \vec{p}_{t1} &= p_t\,(1,0)\,, \\
    \vec{k}_{ti} &= k_{ti}\, (\cos \phi_i , \sin \phi_i) \,,
  \end{split}
\end{equation}
where $\phi_i$ denotes the angle made by the $i^{\mathrm{th}}$
emission with respect to the direction of lepton 1 in the transverse
plane.  It is thus clear that, expressed in these terms, the
transverse component of the $Z$ boson $p_T$ is simply $- \sum_i k_{ti}
\sin \phi_i$ and one has
\begin{equation}
a_T
= \left | \sum_{i} k_{ti} \sin \phi_i \right|\>.
\end{equation}
We note immediately that the dependence on soft emissions is identical
to the case of azimuthal angle $\Delta \phi$ between final state
dijets near the back-to--back region $\Delta \phi \approx \pi$, for
which resummation was carried out in ref.~\cite{BanDasDel}.  This is
not surprising since the component of the $Z$ boson $p_T$, transverse
to the axis defined above, is proportional to $\pi-\Delta\phi$, where
$\Delta \phi$ is the angle between the leptons in the plane transverse
to the beam.  The other (longitudinal) component of $Z$ boson $p_T$,
$a_L$, is proportional in the soft limit to $p_{t1}-p_{t2}$ the
difference in lepton transverse momenta.\footnote{For the case of
  dijet production this $p_t$ imbalance has also been addressed via
  resummation in ref.~\cite{BanDasSymm} which to our knowledge is the
  first extension of the $p_T$ resummation formalism to observables
  involving final state jets.} This kinematics is summarised in
fig.~\ref{fig:at-axis}, which shows final state momenta in the
transverse plane. Together with $\vec p_{t1}$ and $\vec p_{t2}$, the
two lepton transverse momenta, we have displayed the vector boson
transverse momentum $\vec p_T$, the axis $\hat n$ defined in
eq.~(\ref{eq:axis}), and the two transverse momentum components $\vec
a_L$ and $\vec a_T$. From the figure it is also clear that the angle
$\pi-\Delta\phi$, also indicated, is well approximated by $|\vec
a_T|/|\vec p_{t2}|\approx a_T/p_{t1}$. In the case of dijet production
the kinematics is the same, with $\vec p_{t1}$ and $\vec p_{t2}$
representing the transverse momenta of the two highest-$p_t$ jets.

Since it is possible to measure more
accurately the lepton angular separation compared to their $p_t$
imbalance (where momentum resolution is an issue), one can obtain more
accurate measurements of $a_T$ as compared to $a_L$ or the $Z$ boson
$p_T$ which is given by $\sqrt{a_T^2+a_L^2}$~\cite{wyattvester}. The
resummation that we carry out here will be similar in several details
to those of refs.~\cite{BanDasDel,BanDasSymm} but simpler since the
final state hard particles are colourless leptons.  We carry out a
leading order calculation of the observable in question in the
following section and finally the resummation.

\section{Leading order result}
\label{sec:leading}
Here we shall mention how to compute the logarithmically enhanced
terms in $a_T$ at leading order in $\alpha_s$, with details of the
derivation left to the appendix~\ref{sec:Born}. We shall highlight the
origin of the double and single logarithms and relate them to the
corresponding logarithms in the standard $p_T$ distribution at the
same order. The discussion here should facilitate an understanding of
the resummation we carry out in the next section and the results of
subsequent sections.

At Born level we have to consider the process $p_1+p_2 = l_1+l_2$
where $p_1$, $p_2$ and $l_1$, $l_2$ are the four momenta of incoming
partons and outgoing leptons respectively.  The squared matrix element
for the Drell-Yan process for lepton pair production via $Z$ decay
is~\cite{AuLi}
\begin{equation}
  \label{eq:M2DY}
  \mathcal{M}^2_\mathrm{{DY}}(l_1,l_2)= 
  \frac{8}{N_c} \, \mathcal{G}\left(\alpha,\theta_W,M^2,M_Z^2\right) 
  \left [A_l \,A_q\, (t_1^2+t_2^2)+B_l \,B_q\, (t_1^2-t_2^2) \right ],
\end{equation}
with the precise forms of the electroweak constant coefficients
$A_l,A_q$ and $B_l,B_q$ as well as $\mathcal{G}$ reported in
eq.~(\ref{eq:GAB}).  Henceforth we shall suppress the dependence of
$\mathcal{G}$, which has dimension $M^{-4}$, on the standard
electroweak parameters $\alpha,\theta_W,M_Z$. The factor $1/N_c$ comes
from the average over initial state colours.

We have also defined the invariants\footnote{The quantities $t_1$ and
  $t_2$ were labelled as $\hat{t}_1$, $\hat{t}_2$ while $l_1$ and $l_2$
  were labelled $k_1$ and $k_2$ in ref.~\cite{AuLi}.}
\begin{equation}
  t_1 = -2 p_1.l_1 \, \qquad \qquad t_2 = -2 p_2.l_1\,,
\end{equation}
while $M^2$ is the invariant mass of the lepton pair which we fix.
The component $t_1^2+t_2^2$ is the parity conserving piece also
present in the case of the virtual photon process while the
$t_1^2-t_2^2$ component is related to the parity violating piece of
the electroweak coupling and hence absent for the photon case.

We shall study the integrated cross-section which is directly related
to the number of events below some fixed value of $a_T$
\begin{equation}
\label{eq:intcross}
\Sigma(a_T) = \int_0^{a_T} \frac{d^2\sigma}{d a_T'dM^2}\, da_T'\,,
\end{equation}
from which the distribution in $a_T$ can be obtained by
differentiation and the dependence of $\Sigma$ on $M^2$ will be
henceforth implied.

At the Born level the $p_T$ of the lepton pair and hence the $a_T$
vanishes so that the full Born contribution, evaluated at fixed mass
$M^2$ contributes to the cross-section in eq.~\eqref{eq:intcross}.
Evaluating this quantity is straightforward and further explanation is
available in appendix~\ref{sec:Born}. The result we obtain is
\begin{equation}
  \Sigma^{(0)}(a_T) = \mathcal{G} \frac{M^2}{3 \pi}\, \frac{A_l \,A_q}{N_c} 
  \int_0^1 dx_1 \int_0^1 dx_2\,\left [ f_q(x_1) f_{\bar{q}}(x_2) + 
    q \leftrightarrow \bar{q} \right ]\>
  \delta \left (M^2-s x_1 x_2\right)\equiv \Sigma^{(0)}\,,
\end{equation}
where $f_q(x_1)$ and $f_{\bar{q}}(x_2)$ denote parton distribution
functions and $x_1$, $x_2$ the momentum fractions carried by the
incoming partons.  We also do not explicitly indicate a sum over
quark/anti-quark flavours $q$ which should be understood. In writing
the above result we have for simplicity integrated inclusively over
the lepton rapidities, which results in the fact that the parity
violating contribution proportional to $B_l\,B_q$ averages to zero. It
is straightforward to adapt our results to include for instance
acceptance cuts when final experimental data becomes available.

We now derive the QCD corrections to leading order in $\alpha_s$ with
the aim of identifying logarithmically enhanced terms in $a_T$ to the
integrated cross-section defined in eq.~\eqref{eq:intcross}. To this
end we need to consider the process $p_1+p_2 = l_1+l_2+k$ where $k$ is
a final state parton emission as well as $\mathcal{O} \left (\alpha_s
\right)$ virtual corrections to the Drell-Yan process.

Let us focus first on the real emission contribution.  We need to
compute the quantity
\begin{equation}
  \label{eq:sigma1}
  \begin{split}
  \Sigma^{(1)}(a_T) & = 
  \int_0^1 dx_1 \int_0^1 dx_2\,\left\{\left [ f_q(x_1) f_{\bar{q}}(x_2) + 
    q \leftrightarrow \bar{q} \right ]\> \hat \Sigma_A^{(1)}(a_T)\right.\\
  &\left.+\left[ \left(f_q(x_1)+f_{\bar{q}}(x_1)\right) f_g(x_2)+
    q,\bar{q}\leftrightarrow g \right] \hat \Sigma_C^{(1)}(a_T)\right\}\,,
  \end{split}
\end{equation}
where the partonic quantities $\hat{\Sigma}^{(1)}_{A/C}$ which give
the $\mathcal{O}(\alpha_s)$ contribution read
\begin{equation}
  \hat{\Sigma}^{(1)}_i(a_T) = \int d \Phi(l_1,l_2,k) \, \mathcal{M}^2_i (l_1,l_2,k) \delta \left (M^2 -2 l_1.l_2 \right ) \Theta\left(a_T-k_t |\sin \phi| \right),
\end{equation}
where the index $i$ runs over the contributing subprocesses at this
order, i.e.~$i=A/C$ denotes the annihilation (Drell-Yan)/Compton
subprocesses while $\mathcal{M}^2_i$ is the appropriate squared matrix
element the explicit form of which we include in
appendix~\ref{sec:app-lo}. We also need to carry out the integration
over the three-body Lorentz invariant phase space $\Phi$, since in
addition to the final state lepton four-momenta $l_1$, $l_2$ we also
have a final state emitted parton $k$. We have introduced a delta
function constraint that indicates we are working at fixed invariant
mass of the lepton pair $2 l_1.l_2 =M^2$. Additionally in order to
compute the integrated $a_T$ cross-section eq.~\eqref{eq:intcross}, we
need to restrict the additional parton emission $k$ such that we are
studying events below some value of $a_T$. Recalling, from the
previous section, that the value of this quantity generated by a gluon
with transverse momentum $k_t$ and angle with the lepton axis $\phi$
is $k_t |\sin \phi|$ we arrive at the step function in the above
equation.\footnote{As we stated previously this approximation is
  sufficient up to terms that vanish as $a_T \to 0$, which we do not
  compute here.}  We then fold the parton level result with parton
distribution functions precisely as for the Born level result
$\Sigma^{(0)}$ reported above.

After integrating over all lepton variables, accounting for virtual
corrections and retaining only singular terms in the limit $k_t \to 0$
(which are the source of logarithms in $a_T$), as detailed in
appendix~\ref{sec:Born}, we arrive at the result for the annihilation
contribution
\begin{multline}
\label{eq:ann-sigma1}
\Sigma^{(1)}_A(a_T) = -\mathcal{G} \frac{M^2}{3\pi} \frac{A_lA_q}{N_c}
\int_0^1 \frac{d\Delta}{\Delta} \int_0^{1-2\sqrt{\Delta}} \!\!\!\!\!
dz\,\, \times \\ \times \int_0^1 dx_1 \int_0^1 dx_2
\left[f_q(x_1)f_{\bar{q}}(x_2)+q\leftrightarrow \bar{q} \right ]\, 
\delta \left(M^2-s x_1 x_2 z\right) \times \\
\times \int_0^{2 \pi}\frac{d\phi}{2\pi}\, C_F
\frac{\alpha_s}{2\pi}\frac{2\left(1+z^2\right)}{\sqrt{(1-z)^2-4z
    \Delta}} \Theta \left(\sqrt{\Delta} |\sin \phi| -\frac{a_T}{M}
\right),
\end{multline}
while that for the Compton subprocess reads 
\begin{multline}
\label{eq:comp-sigma1}
\Sigma^{(1)}_C(a_T) = -\mathcal{G}\frac{M^2}{3\pi} \frac{A_l A_q}{N_c}
\int_0^1 \frac{d\Delta}{\Delta}
\int_0^{1-2\sqrt{\Delta}}\!\!\!\!\! dz\,\,
 \times \\ \times 
\int_0^1 \!\!dx_1 \int_0^1 \!\!dx_2 
\left[
  \left(f_q(x_1)+f_{\bar{q}}(x_1)\right)
  f_g(x_2)+q,\bar{q}\leftrightarrow g \right] \,
\delta
\left(M^2- s x_1 x_2 z\right)
 \times \\ \times
\int_0^{2 \pi} \frac{d\phi}{2\pi}
\, T_R \frac{\alpha_s}{2\pi}
\frac{(1-z)\left(1-2z+2 z^2\right)}{\sqrt{(1-z)^2-4 z \Delta}} \Theta
\left(\sqrt{\Delta} |\sin \phi| -\frac{a_T}{M} \right).
\end{multline}
Note that the above equations involve the step function constraint
$\Theta \left ( k_t |\sin \phi | -a_T \right)$ which represents the
fact that the number of events with $k_t |\sin \phi| < a_T$ is equal
to the total rate minus the events with $k_t |\sin \phi| >a_T$. Since
the total rate is a number independent of $a_T$, we can simply compute
the events with $k_t |\sin \phi| > a_T$ to obtain the logarithmic
$a_T$ dependence, which is what we have done above.

In the above equations we have also parametrised the integral over the
gluon momentum via the rescaled transverse momentum $\Delta =
k_t^2/M^2$, the azimuthal angle $\phi$ and $z$ where in the collinear
limit $1-z$ is just the fraction of the parent partons energy carried
off by the radiated gluon.  We also have as usual $C_F=4/3$,
$T_R=1/2$, $\alpha_s =g^2/4 \pi$.

The above results are sufficient to obtain the logarithmic structure
in $a_T$ and compare it to the corresponding result for the $Z$ boson
$p_T$ distribution.  In this respect we note that the only difference
between the results reported immediately above and those for the $p_T$
case are the $|\sin \phi|$ terms in the step function constraints
above. While at the leading order these will essentially just be a
matter of detail we shall see that the $\sin \phi$ dependence has an
important role to play in the shape of the resummed spectrum.

To complete the calculations one proceeds as in the $Z$ boson $p_T$
case and hence we take the moments with respect to the standard
Drell-Yan variable $\tau = \frac{M^2}{s}$, thereby defining
\begin{equation}
\tilde{\Sigma}(N,a_T) = \int_0^{1} d\tau \,\tau^{N-1} \,\Sigma(a_T),
\end{equation}
which can be expressed as a sum over the moment space annihilation and 
Compton terms $\tilde{\Sigma}(N,a_T) = \tilde{\Sigma}_A(N,a_T)+
\tilde{\Sigma}_C \left(N,a_T \right)$. 

The Born level Drell-Yan contribution can then be expressed in moment
space as
\begin{equation}
\label{eq:born}
\tilde{\Sigma}^{(0)}(N) = \frac{\mathcal{G}}{3\pi} \frac{A_l A_q}{N_c} \,F_A(N)\,,
\end{equation}
where $F_A(N)$ denotes the moment integrals of the parton distribution
functions
\begin{equation}
\label{eq:pdf-N}
\begin{split}
  F_A(N)&= \int_0^1 \,dx_1 \,x_1^{N} \int_0^1 dx_2\, x_2^{N} \,\left
    [f_q(x_1)\,f_{\bar{q}}(x_2)+ q \leftrightarrow \bar{q} \right ]\\
  &=\tilde f_q(N)\,\tilde f_{\bar{q}}(N)+ q \leftrightarrow \bar{q}
  \,,
\end{split}
\end{equation}
where we introduced $\tilde{f}(N)$, the moments of the parton
distributions.

Likewise the $\mathcal{O}(\alpha_s)$ annihilation contribution can be
expressed as
\begin{multline}
  \tilde{\Sigma}^{(1)}_A(N,a_T) = -\frac{\mathcal{G}}{3 \pi}\,\frac{A_l A_q}{N_c}
  F_A(N)\int_0^1 \frac{d\Delta}{\Delta} \int_0^{1-2\sqrt{\Delta}}\!dz \,z^{N}  \times \\
  \times  \int_0^{2\pi}\frac{d\phi}{2\pi} \, C_F \frac{\alpha_s}{2\pi}
  \frac{2\left(1+z^2\right)}{\sqrt{(1-z)^2-4z \Delta}} \Theta
  \left(\sqrt{\Delta} |\sin \phi| -\epsilon \right)\,,
\end{multline}
where $\epsilon = a_T/M$ is a dimensionless version of the $a_T$
variable.

Performing the integrals over $z$ and $\Delta$ we obtain the result
\begin{multline}
  \tilde{\Sigma}^{(1)}_A(N,a_T) = -\tilde{\Sigma}^{(0)}(N)\left[
    2\frac{\alpha_s}{\pi}\gamma_{qq}(N) \int_0^{2\pi}\!\frac{d \phi}{2\pi}
    \ln \frac{|\sin\phi|}{\epsilon} \right.\\
  \left.+\frac{2 C_F \alpha_s}{\pi}
    \int_0^{2\pi}\!\frac{d\phi}{2\pi} \left ( \ln^2 \frac{|\sin
        \phi|}{\epsilon}-\frac{3}{2} \ln \frac{|\sin
        \phi|}{\epsilon}\right)\right]\,.
\end{multline}
where we introduced the quark anomalous dimension
\begin{equation}
\gamma_{qq}(N) =
C_F \int_0^1 dz \left(z^{N}-1\right) \frac{1+z^2}{1-z}\,.
\end{equation}
Notice the proportionality of the above result to the Born level
result which is a consequence of the collinear origin of logarithmic
terms.

We have not integrated over the variable $\phi$ as yet in order to
make the link to results for the $p_T$ distribution. To obtain the
$\mathcal{O}(\alpha_s)$ integrated cross-section for the $p_T$ case
the same formulae as reported above apply but one replaces $|\sin
\phi|$ by unity while $\epsilon$ would denote $p_T/M$. The $\phi$
integral is then trivial and can be replaced by unity.  For the $a_T$
variable on performing the $\phi$ integral we use the results
\begin{eqnarray}
  \int_0^{2\pi} \frac{d\phi}{2\pi} \ln^2|\sin \phi| &=& \ln^2 2+\frac{\pi^2}{12}, \\
  \int_0^{2\pi} \frac{d \phi}{2 \pi} \ln |\sin \phi| &=& -\ln 2,
\end{eqnarray} 
to obtain
\begin{multline}
    \label{eq:FAN}
  \tilde{\Sigma}^{(1)}_A(N,a_T) = -\,\tilde{\Sigma}^{(0)}(N)\times \\
  \times \left[2\frac{\alpha_s}{\pi} 
    \gamma_{qq}(N) \ln \frac{1}{2\epsilon}+\frac{2 C_F \alpha_s}{ \pi} 
    \left ( \ln^2 \frac{1}{2 \epsilon}-\frac{3}{2} 
      \ln \frac{1}{2\epsilon} \right)
    +\frac{C_F \alpha_s}{2\pi} \frac{\pi^2}{3}\right]\,.
\end{multline}

The corresponding result for the QCD Compton process is purely single
logarithmic and reads
\begin{equation}
  \begin{split}
    \label{eq:FCN}
    \tilde{\Sigma}^{(1)}_C(N,a_T) &= -\frac{\mathcal{G}}{3 \pi}
    \frac{A_l A_q}{N_c} F_C(N)
    \left(2\frac{\alpha_s}{\pi}\gamma_{qg}(N)
      \int_0^{2\pi}\!\frac{d\phi}{2\pi}\ln \frac{|\sin
        \phi|}{\epsilon}
    \right) \\
    &= -\frac{\mathcal{G}}{3 \pi} \frac{A_l A_q}{N_c} F_C(N)\,
    2\frac{\alpha_s}{\pi} \gamma_{qg}(N) \ln \frac{1}{2 \epsilon}\,,
  \end{split}
\end{equation}
where
\begin{equation}
  \gamma_{qg}(N) = T_R \int_0^1\!dz\, z^{N} \left[z^2+(1-z)^2\right]\,,
\end{equation}
and $F_C(N)$ is the moment integral of the relevant combination of
parton density functions 
\begin{equation}
  \begin{split}
  F_C(N) &= 
  \int_0^1 \,dx_1 \,x_1^{N} \int_0^1 dx_2\, x_2^{N} \,
  \left[
    \left(f_q(x_1)+f_{\bar{q}}(x_1)\right)
    f_g(x_2)+q,\bar{q}\leftrightarrow g \right]\\
  &= \left(\tilde f_q(N)+\tilde f_{\bar{q}}(N)\right)
    \tilde f_g(N)+q,\bar{q}\leftrightarrow g\,.
  \end{split}
\end{equation}
In our final results, eqs.~(\ref{eq:FAN}) and (\ref{eq:FCN}), we have
neglected constant terms that are identical to those for the Drell-Yan
$p_T$ distribution computed for instance in~\cite{GrazziniDeFlorian}.

We note that the logarithms found here, both in the Drell-Yan and
Compton contributions, are the same as those for the $p_T$ variable
with the replacement $\epsilon \to 2 \epsilon$. In other words as far
as the logarithmic dependence is concerned we obtain that the result
for the cross-section for events with $a_T < \epsilon M$ is the same
as the result for the variable $p_T/2 < \epsilon M$. The only other
effect, at this order, of the $|\sin \phi|$ term is to generate a
constant term $\frac{C_F \alpha_s}{2\pi} \frac{\pi^2}{3}$ reported
above.  Thus to leading order in $\alpha_s$ we have simply
\begin{equation}
\label{eq:lodiff}
\Sigma^{(1)}(a_T)|_{a_T=\epsilon M}-\Sigma^{(1)}
\left (\frac{p_T}{2}\right)|_{p_T/2=\epsilon M }= 
-\Sigma^{(0)}\,C_F \frac{\alpha_s}{2\pi} \frac{\pi^2}{3}.
\end{equation}
In writing the above we have returned to $\tau$ space by inverting the
Mellin transform so as to obtain the result in terms of the factor
$\Sigma^{(0)}$ rather than $\tilde{\Sigma}^{(0)}(N)$.

The result above can be verified by using a fixed-order program such
as MCFM.  One can obtain the results for the integrated cross-sections
for $a_T$ and $p_T/2$ and the difference between them should be a
constant with the value reported above. This is indeed the case, as
one can see from the plot in figure~\ref{fig:at-vs-pt}, where the
difference in eq.~(\ref{eq:lodiff}) generated using the numerical
fixed-order program MCFM~\cite{CampEll}, divided by the Born cross
section $\Sigma^{(0)}$, is plotted against $L=\ln(\epsilon)$. The
results from MCFM agree with our expectation \eqref{eq:lodiff}. In
order to show the smoothest curve we have taken the case where the $Z$
decay has been treated fully inclusively (i.e.\ we have not placed
rapidity cuts) and a narrow width approximation eventually employed
but we have checked our results agree with MCFM for arbitrary cuts on
lepton rapidities.  \FIGURE{
  \epsfig{file=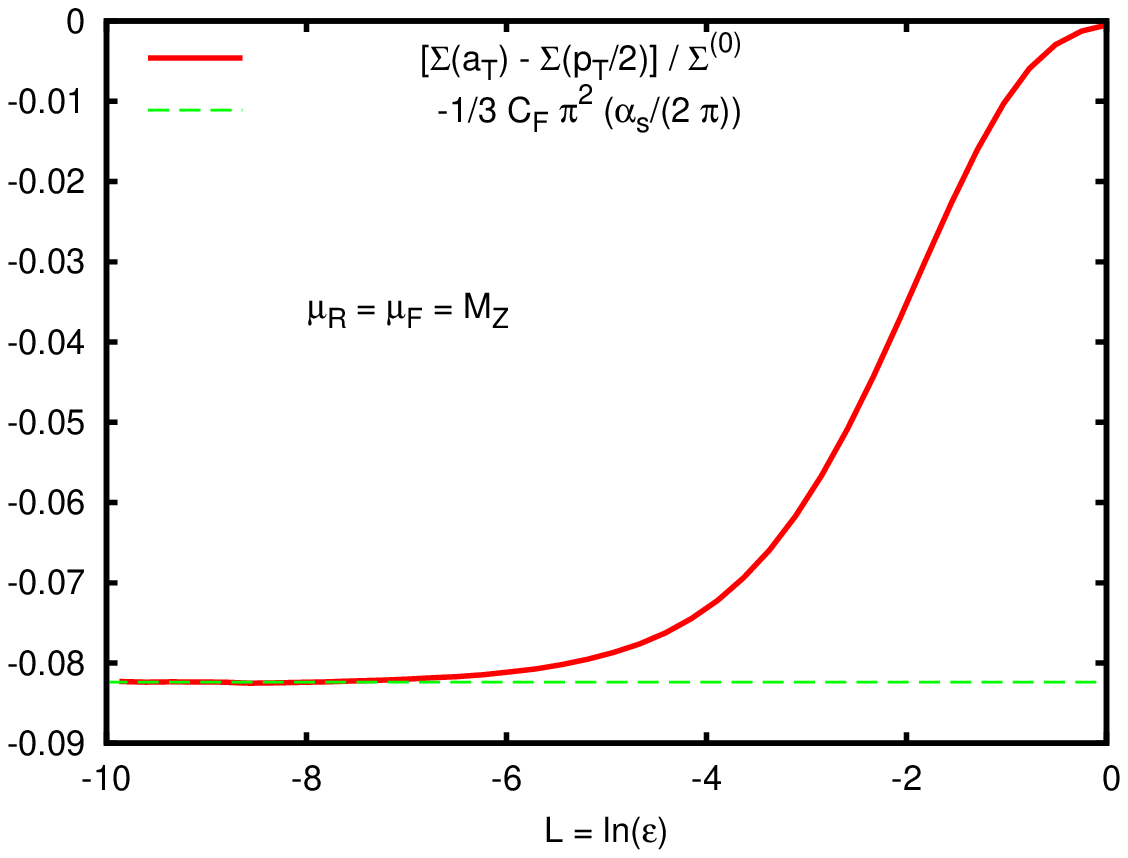,width=.9\textwidth}
  \caption{The difference between the integrated distributions for
    $a_T$ and $p_T/2$. Here we have used the CTEQ6M pdf set~\cite{cteq}
    and both factorisation and renormalisation scales $\mu_F$ and
    $\mu_R$ have been fixed at the $Z$ boson mass $M_Z$.}
  \label{fig:at-vs-pt}
}
Having carried out the fixed-order computation, which serves to
illustrate some important points, we shall shift our attention to the
resummation of logarithms to all orders. 

\section{Resummed Results}
Here we shall carry out the resummation of the large logarithms in the
ratio of two scales $M$ and $a_T$ which become disparate at small
$a_T$, $a_T \ll M$.  We already derived the dependence of the $a_T$ on
multiple soft and/or collinear emissions in the preceding section and
hence in order to carry out the resummation we next need to address
the dynamics of multiple low $k_t$ emissions. We shall first treat
only the Drell-Yan process and later specify the role of the QCD
Compton production process.

We shall study as before the integrated cross-section representing the
number of events below some fixed value of $a_T$, defined in
eq.~(\ref{eq:intcross}), from which one can obtain the $a_T$
distribution by differentiating with respect to $a_T$.  Also as we
emphasised in the previous section we are working at fixed invariant
mass of the lepton pair purely as an illustrative example and we can
straightforwardly adapt our calculations to take into account
experimental cuts on for instance lepton rapidities, which in any case
do not affect the resummation.

We consider again the incoming partons as carrying momentum fractions
$x_1$ and $x_2$ of the incoming hadrons which means that at Born level
where they annihilate to form the lepton pair via virtual $Z$
production we have simply $M^2 = \hat{s} = s x_1 x_2$, where the
Mandelstam invariant $\hat{s}$ denotes the partonic centre of mass
energy. Beyond the Born level one has to take account of gluon
radiation and to this end we introduce as in the previous section the
quantity $z= M^2/\hat{s}$ such that $1-z$ represents the fractional
energy loss of the incoming partons due to the radiation of collinear
gluons. Thus in the limit $z\to 1$ one is probing soft and collinear
radiation while away from $z=1$ we will be dealing with energetic
collinear emission.  We note here that for the purpose of generating
the logarithms we resum we do not have to examine large-angle
radiation and the collinear limit is sufficient as for the usual $p_T$
distribution. In fact since the $a_T$ resummation we aim to carry out
shares several common features with the well-known $p_T$ distribution
we shall only sketch the resummation concentrating instead on features
of the $a_T$ which lead to differences from the $p_T$ variable. For a
recent detailed justification of the approximations that lead to NLL
resummation for the $p_T$ case as well as for other variables the
reader is referred to ref.~\cite{caesar}.

We work in the centre-of--mass frame of the colliding partons and in
moment space where we take moments with respect to $\tau = M^2/s$ of
the cross-section in eq.~(\ref{eq:intcross}) as in the fixed-order
calculations we carried out. Taking moments enables us to write for
the emission of multiple collinear and optionally soft gluons
\begin{equation}
\tilde{\Sigma}(N,a_T) = \tilde{\Sigma}^{(0)}(N)\, W_N(a_T)\,,
\end{equation}
where $\tilde{\Sigma}^{(0)}(N)$ is the Born level result in
eq.~\eqref{eq:born}.  The effects of multiple collinear (and
optionally soft) gluon emission from the incoming projectiles are
included in the function $W_N$ which can be expressed to
next-to--leading logarithmic (NLL) in the standard factorised form
\begin{multline}
\label{eq:prob}
W^{\mathrm{real}}_N(a_T) = \sum_{n=0}^{\infty}
\frac{1}{n!}\prod_{i=1}^{n}
\int dz_i \frac{dk_{t_i}^2}{k_{ti}^2} \frac{d\phi_i}{2\pi} \,\times \\
\times\,
 z_i^{N} \,2 C_F \, \frac{\alpha_s(k_{ti}^2)}{2 \pi}
\left(\frac{1+z_i^2}{1-z_i}\right) 
\Theta \left(a_T-|\sum_i  v(k_i)|\right)\,,
\end{multline}
where $1-z_i$ denotes the fraction of momentum carried away by
emission of a quasi-collinear gluon $i$ from the incoming hard
projectile so that $M^2/\hat{s}= z = \prod_i z_i$ and $k_{ti}$ is the
transverse momentum of gluon $i$ with respect to the hard emitting
incoming partons.

In writing the above results we have used an independent emission
approximation, valid to NLL accuracy, where the emission probability
for $n$ collinear gluons is merely the product of single gluon
emission probabilities, which factorise from the Born level production
of the hard lepton pair.\footnote{This approximation is invalid for
  situations when one is examining soft radiation in a limited angular
  interval away from hard emitting particles~\cite{Dassal1,Dassal2},
  which is not the case here.} The single gluon emission probability
to the same NLL accuracy is given by the leading order splitting
function for the splitting of a quark to a quasi-collinear quark and
gluon (weighted by the running strong coupling),
\begin{equation}
 P_{qq}(z)\frac{\alpha_s(k_{t}^2)}{2\pi} = C_F
\frac{\alpha_s(k_{t}^2)}{2\pi}\frac{1+z^2}{1-z}\,, 
\end{equation}
with $\alpha_s$ defined in the CMW scheme~\cite{CMW}. We have inserted
a factor of two to take account of the fact that there are two hard
incoming partons which independently emit collinear gluons.  We have
also taken care of the constraint on real gluon emission, imposed by
the requirement that the sum of the components of the $k_{ti}$ normal
to the axis in eq.~(\ref{eq:axis}) (denoted by $v(k_i)=k_{ti}
\sin\phi_i$) is less than $a_T$.  We have integrated over the leptons,
holding the invariant mass $M$ fixed, and taken moments to obtain the
full zeroth order Drell-Yan result $\tilde{\Sigma}^{(0)}(N)$, which
multiplies the function $W_N$ containing all-order radiative effects.

All of the above arguments would also apply to the case of the $p_T$
variable. Thus while the dynamics of multiple soft/collinear emission
is treated exactly as for the $p_T$ resummation the difference between
the $p_T$ and our resummation arises purely due to the different form
of the argument of the step function restricting multiple real
emission. Thus while for the $p_T$ variable the phase space constraint
involves a two-dimensional vector sum $\Theta \left (p_T-|\sum_i
  \vec{k_{ti}}| \right)$, in the present case we have a one
dimensional sum of the components of the gluon $k_t$ normal to the
lepton thrust axis, $v(k_i) =k_{ti} \sin \phi_i$.  One encounters such
a one dimensional sum also in cases such as azimuthal correlations in
DIS~\cite{BMS,BanDasDel} and the resummation of the $p_t$ difference
between jets in dijet production~\cite{BanDasSymm}. It is this
difference that will be responsible for different features of the
$a_T$ distribution as we shall further clarify below. The relationship
between azimuthal correlations and the $a_T$ is no surprise since the
$a_T$ variable is proportional to $\pi -\Delta \phi_{ll}$, the
deviation of the azimuthal interval between the leptons from its Born
value $\pi$.

In order to further simplify eq.~\eqref{eq:prob} we also factorise the
phase space constraint using a Fourier representation of the step
function~\cite{BMS}
\begin{equation}
\Theta \left(a_T-|\sum_i v(k_i)|\ \right) = 
\frac{1}{\pi}\int_{-\infty}^{\infty}
\frac{db}{b} \sin(b a_T) \prod_{i} e^{ibv(k_i)}.
\end{equation}
Note the presence of the $\sin(b a_T)$ function which is a consequence
of addressing a one dimensional sum as opposed to the Bessel function
$J_1$ one encounters in resummation of the $p_T$ cross-section.  With
both the multiple emission probability and phase space factorised as
above it is easy to carry out the infinite sum in eq.~\eqref{eq:prob}
which yields
\begin{equation}
W_N^{\mathrm{real}}(a_T) = \frac{1}{\pi}\int_{- \infty}^{\infty} 
\frac{db}{b} \sin(b a_T) \,e^{R^{\mathrm{real}} (b)}\,,
\end{equation}
with the exponentiated real gluon emission contribution
\begin{equation}
  R^{\mathrm{real}(b)} = \int dz \frac{d k_t^2}{k_t^2} \frac{d\phi}{2 \pi} \,
  z^{N}\, 2\, C_F \frac{\alpha_s(k_t^2)}{2 \pi}\left ( \frac{1+z^2}{1-z} \right)
  \,e^{i b v(k)}\, \Theta \left (1-z-\frac{k_t}{M}\right)\,.
\end{equation}
The kinematic limit on the $z$ integration is set in such a way that
one correctly accounts for soft large angle emissions.

Next we include all-order virtual corrections which straightforwardly
exponentiate in the soft-collinear limit to yield finally
\begin{equation}
W_N(a_T) = \frac{2}{\pi}\int_{0}^{\infty} \frac{db}{b} \sin(b\, a_T)\, 
e^{-R(b)}\,,
\end{equation}
where 
\begin{multline}
  -R(b) = R^{\mathrm{real}}+R^{\mathrm{virtual}}= \int dz \frac{d
    k_t^2}{k_t^2} \frac{d\phi}{2 \pi} \, 2 C_F
  \frac{\alpha_s(k_t^2)}{2 \pi} \left (\frac{1+z^2}{1-z} \right)
  \,\times \\ \times\, \left(z^{N}\,e^{i b v(k)}-1\right) \Theta \left
    (1-z-\frac{k_t}{M} \right),
\end{multline}
where it should be clear that the term corresponding to the $-1$ added
to the real contribution $z^{N}e^{i b v(k)}$ corresponds to the
virtual corrections. Note that the virtual corrections are naturally
independent of both Fourier and Mellin variables $b$ and $N$
respectively since they do not change the longitudinal or transverse
momentum of the incoming partons and hence exponentiate directly.  We
are thus left to analyse $R(b)$ the ``radiator'' up to
single-logarithmic accuracy.

\subsection{The resummed exponent}
Here we shall evaluate the function $R(b)$ representing the resummed
exponent to the required accuracy. We shall first explicitly introduce
a factorisation scale $Q_0^2$ to render the integrals over $k_t$
finite. Later we will be able to take the $Q_0 \to 0$ limit. Thus one
considers all emissions with transverse momenta below $Q_0$ to be
included in the pdfs which are defined at scale $Q_0$ such that the
factor $\tilde{\Sigma}^{(0)}(N)$ reads
\begin{equation}
  \tilde{\Sigma}^{(0)}(N) = \frac{A_l A_q}{N_c} \frac{\mathcal{G}}{3\pi} F_A(N,Q_0^2)\,,
\end{equation}
with 
\begin{equation}
  \begin{split}
    F_A(N,Q_0^2) &= \int_0^1 \,dx_1 x_1^{N} \int_0^1 dx_2 x_2^{N} \left
      [f_q(x_1,Q_0^2)f_{\bar{q}}(x_2,Q_0^2)+
      q \leftrightarrow \bar{q} \right ] \\
    &= \left[\tilde{f}_q(N,Q_0^2)\,\tilde{f}_{\bar{q}}(N,Q_0^2)+ q
      \leftrightarrow \bar{q} \right ]\,.
  \end{split}
\end{equation}
Thus while in the fixed-order calculation of the previous section the
pdfs could be treated as bare scale independent quantities for the
resummed calculation we start with full pdfs evaluated at an arbitrary
(perturbative) factorisation scale. The $k_t$ integration in the
perturbative radiator should now be performed with scale $k_t > Q_0$.
We next follow the method of ref.~\cite{Dassalthrust} where an
essentially identical integral was performed for the radiator.

First, following the method of ref.~\cite{Dassalthrust} we change the
argument of the pdfs from $Q_0$ to the correct hard scale of the
problem, the pair invariant mass $M$, via DGLAP evolution.  To be
precise we use for the quark distribution
\begin{equation}
  \tilde{f}_q(N,Q_0^2) =\tilde f_{\bar{q}}(N,M^2)\> 
  e^{-\int_{Q_0^2}^{M^2} \frac{dk_t^2}{k_t^2} \frac{\alpha_s(k_t^2)}{2 \pi} \gamma_{qq} (N)}\,,
\end{equation}
and likewise for the anti-quark distribution where $\gamma_{qq}(N)$ is
the standard quark anomalous dimension matrix. Note that we have not
yet considered the QCD Compton scattering process and the
corresponding evolution of the quark pdf from incoming gluons via the
$\gamma_{qg}$ anomalous dimension matrix, which we shall include in
the final result by using the full pdf evolution rather than the
simplified form reported immediately above.  Carrying out the above
step results in a modified radiator such that one now has
\begin{multline}
  R(b) = 2 C_F  \int_{Q_0^2}^{M^2} \frac{dk_t^2}{k_t^2} 
  \frac{\alpha_s(k_t^2)}{2\pi}
  \int_0^{2 \pi}\frac{d\phi}{2\pi} \,\times \\ 
  \times \,\left [\int_0^1 \!dz\, \frac{1+z^2}{1-z} 
    \left(1-z^{N} e^{ibv(k)}\right)\Theta \left(1-z-\frac{k_t}{M}\right)+
    \frac{\gamma_{qq}(N)}{C_F} \right ].
\end{multline}
Using the definition of the anomalous dimension $\gamma_{qq}(N)$ we
can write the above (see for instance ref.~\cite{Dassalthrust}) as
\begin{equation}
  R(b) = 2 C_F \int_{Q_0^2}^{M^2}\frac{dk_t^2}{k_t^2} \frac{\alpha_s(k_t^2)}{2\pi}
  \int_0^{2 \pi}\frac{d\phi}{2\pi} 
  \int_0^1 dz \,z^{N} \frac{1+z^2}{1-z} 
  \left (1-e^{ibv(k)}\right )\Theta \left(1-z-\frac{k_t}{M}\right)\,,
\end{equation}
where in arriving at the last equation we neglected terms of $\mathcal{O}\left(k_t/M \right)$.

To NLL accuracy we can further make the approximation~\cite{BMS}
\begin{equation}
1-e^{ibv(k)} \approx  \Theta \left (k_t |\sin\phi| - 
\frac{1}{\bar{b}}\right)\,,
\end{equation}
where we used the fact that in the present case $v(k_i) = k_t \sin
\phi_i$ with $\bar{b}= b \,e^{\gamma_E}$.  Thus one gets for the radiator
$R(b) \equiv R(\bar{b})$ with
\begin{multline}
  R(\bar{b}) =2 C_F\int_{Q_0^2}^{M^2}\frac{dk_t^2}{k_t^2} 
  \frac{\alpha_s(k_t^2)}{2\pi}
  \int_0^{2 \pi}\frac{d\phi}{2\pi} \int_0^1 dz \,z^{N} \frac{1+z^2}{1-z}
  \times \\ \times
  \Theta\left( k_t|\sin\phi|-\frac{1}{\bar{b}} \right ) 
  \Theta \left(1-z-\frac{k_t}{M}\right)\,.
\end{multline}
We now evaluate the radiator to the required accuracy. Performing the
$z$ integration to NLL accuracy, and neglecting terms of relative
order $k_t/M$, one arrives at
\begin{equation}
\label{eq:radfull}
R(\bar{b}) = 2 C_F \int_0^{2 \pi} \frac{d\phi}{2\pi} 
\int_{0}^{M^2} \frac{dk_t^2}{k_t^2}\frac{\alpha_s(k_t^2)}{2 \pi}
\Theta \left(k_t |\sin \phi|-\frac{1}{\bar{b}}\right) 
\left (\ln \frac{M^2}{k_t^2} -\frac{3}{2} +
  \frac{\gamma_{qq}(N)}{C_F}\right )\,,
\end{equation}
where in the last line we took $Q_0\! \to\! 0$ since the $k_t$
integral is now cut-off by the step function $\Theta \left[k_t|\sin
  \phi| -1/\bar{b}\right]$ and chose to perform the $\phi$ integration
at the end.  We note that the only difference between the radiator
above and the standard $p_T$ distribution is the factor $|\sin \phi|$
multiplying $k_t$ in the step function condition above.  This will
result in an additional single-logarithmic contribution not present in
the $p_T$ resummation results.

In order to deal with the $\phi$ dependence to NLL accuracy we expand
(as in ref.~\cite{BanDasSymm}) eq.~(\ref{eq:radfull}) about $|\sin
\phi|=1$ in powers of $\ln |\sin \phi|$ to obtain
\begin{multline}
\label{eq:radnll}
R(\bar{b}) = 2 C_F \int_{0}^{M^2}
\frac{dk_t^2}{k_t^2}\frac{\alpha_s(k_t^2)}{2 \pi} \Theta \left(k_t
  -\frac{1}{\bar{b}}\right)
\left (\ln \frac{M^2}{k_t^2} -\frac{3}{2} +\frac{\gamma_{qq}(N)}{C_F}\right ) \\
+ \frac{\partial R(\bar b)}{\partial \ln (\bar b M)} \, \int_0^{2\pi}
\frac{d\phi}{2\pi}\,\ln |\sin \phi| +\cdots
\end{multline}
where we have neglected higher derivatives of $R$ as they will
contribute only beyond NLL accuracy. Moreover in evaluating $\partial
R(\bar b)/\partial \ln(\bar b M)$ we can replace $R$ by its leading
logarithmic form $R_{\mathrm{LL}} (\bar{b})$ since logarithmic
derivatives of any next-to-leading logarithmic pieces of $R(\bar{b})$
will give only NNLL terms that are beyond our accuracy.  The first
term on the RHS of the above equation is in fact just the radiator we
would get for resummation of the $Z$ boson $p_T$ distribution which
contains both leading and next-to-leading logarithmic terms. The
second term on the RHS accounts for the $\phi$ dependence of the
problem and is purely next-to--leading logarithmic in nature since it
contains the logarithmic derivative of $R_{\mathrm{LL}}(\bar{b})$.
Thus we need to evaluate the first term on the RHS of
eq.~(\ref{eq:radfull}) and then isolate its leading-logarithmic piece
to compute the second term on the RHS.

We carry out the integral over $k_t$ to NLL accuracy using standard
techniques~\cite{CTTW}, i.e.~we change the coupling from the CMW to
the $\bar{\mathrm{MS}}$ scheme
\begin{equation}
  \alpha^\mathrm{CMW}_s(k_t^2)= \alpha_s^{\bar{\mathrm{MS}}}(k_t^2)
  \left(1+K \,\frac{\alpha_s^{\bar{\mathrm{MS}}}(k_t^2)}{2\pi}\right)\,,\qquad
  K=C_A \left(\frac{67}{18}-\frac{\pi^2}{6} \right)-\frac{5}{9} n_f\,,
\end{equation}
and use a two-loop running coupling 
\begin{equation}
  \label{eq:twoloop-as}
  \alpha_s(k_t^2) = 
  \frac{\alpha_s(M^2)}{1-\rho}\left[1-\alpha_s(M^2)\frac{\beta_1}{\beta_0}
    \frac{\ln(1-\rho)}{1-\rho}\right] \,,\qquad
  \rho = \alpha_s(M^2) \beta_0 \ln\frac{M^2}{k_t^2}\,,
\end{equation}
where the beta function coefficients are defined as
\begin{equation}
\beta_0 = \frac{11 C_A-2 n_f}{12 \pi}, \, \, \qquad \beta_1 =\frac{17 C_A^2-5C_A n_f -3 C_F n_f}{24 \pi^2}.
\end{equation}
and $\alpha_s(M^2)$ is a shorthand for
$\alpha_s^{\bar{\mathrm{MS}}}(M^2)$. 
We then obtain the usual expression
\begin{equation}
R(\bar{b})= L g_1(\alpha_s L) + g_2 (\alpha_s L)\,,
\end{equation}
with $L \equiv \ln\left (\bar{b}^2 M^2 \right)$. The functions $g_1$
and $g_2$ are then the leading and next-to--leading logarithmic
functions which have the following detailed form
\begin{align}
  g_1(\lambda) =& \frac{C_F}{\pi\beta_0\lambda} \left
    [-\lambda-\ln{(1-\lambda)}\right]\,,\\
  g_{2}({\lambda})=&  \frac{3C_F}{2\pi \beta_0} \ln(1-\lambda)-\frac{\gamma_{qq}(N)}{\pi\beta_0} \ln(1-\lambda)+\frac{2 C_F}{\pi \beta_0} \frac{\lambda}{1-\lambda}(-\ln 2) \nonumber \\
  +&\frac{KC_F [\lambda + (1-\lambda)\ln(1-\lambda)]}
  {2\pi^2\beta_0^2(1-\lambda)} -\frac{C_F \beta_1}{\pi\beta_0^3}
  \left[ \frac{\lambda + \ln (1-\lambda)}{1-\lambda} + \frac{1}{2}
    \ln^2{(1-\lambda)} \right],
\label{eq:g2}
\end{align}
with $\lambda =  \alpha_s(M^2)\,\beta_0 L$.

Let us comment on the origin of various terms. The leading logarithmic
function $g_1(\lambda)$ arises from soft and collinear emission
integrated over the phase space with a running coupling (to be precise
the one-loop running of the coupling is sufficient to give us $g_1$).
It is identical to the corresponding function for $Z$ boson $p_T$
resummation and at this level the $a_T$ and $p_T$ variables do not
differ. The function $g_2$ embodies hard-collinear radiation (and
hence the appearance of the quark anomalous dimension
$\gamma_{qq}(N)$) as well as the two-loop running of the coupling and
the change to the $\bar{\mathrm{MS}}$ scheme from the CMW scheme which
gives rise to the piece proportional to $K$.  It is also the same as
the corresponding function for $Z$ boson $p_T$ resummation
\emph{except} for the additional single-logarithmic term
$\frac{2C_F}{\pi \beta_0} \frac{\lambda}{1-\lambda}(-\ln 2)$ which
arises from the $\phi$ dependence of the problem.  In other words one
has explicitly
\begin{equation}
  \frac{\partial R_{\mathrm{LL}}(\bar{b})}{\partial \ln (\bar{b}M)} 
  \int_0^{2 \pi} \frac{d\phi}{2\pi}\ln |\sin \phi| = 
  \frac{\partial L g_1(\lambda)}{\partial \ln (\bar{b}M)}(-\ln 2)=  
  \frac{2C_F}{\pi \beta_0} \frac{\lambda}{1-\lambda}(-\ln 2)\,.
\label{eq:Rapprox2}
\end{equation}
However, this term, within NLL accuracy, can be absorbed in the
radiator with a change in the definition of its argument $\bar b$:
\begin{equation}
  \label{eq:rad-newbarb}
  R(\bar b) +
  \frac{\partial R_{\mathrm{LL}}(\bar{b})}{\partial \ln (\bar{b}M)}   
  \int_0^{2 \pi} \frac{d\phi}{2\pi}\ln |\sin \phi| =
  R(\bar b) -
  \frac{\partial R_{\mathrm{LL}}(\bar{b})}{\partial \ln (\bar{b}M)} \ln 2 
  \simeq 
  R(\bar b/2)\,,
\end{equation}
and $R(\bar b/2) = R(b \,e^{\gamma_E}/2)$ is precisely the radiator
for the Z boson $p_T$ distribution (see e.g. ref.~\cite{Catani}).  As
a final step we can use the anomalous dimension matrix
$\gamma_{qq}(N)$ and the corresponding contribution $\gamma_{qg}(N)$
from Compton scattering which we have for brevity avoided treating, to
evolve the pdfs from scale $M^2$ to scale $(2/\bar{b})^2$ precisely as
in the case of the $p_T$ variable. After absorption of the $N$
dependent piece of the radiator into a change of scale of the pdfs it
is trivial to invert the Mellin transform to go from $N$ space to
$\tau$ space. We can thus schematically write the result for the $a_T$
cross-section defined in \eqref{eq:intcross} resummed to NLL accuracy
as
\begin{multline}
\label{eq:resummed}
\Sigma(a_T) = \mathcal{G} \frac{M^2}{3\pi} \frac{A_l A_q}{N_c} 
\frac{2}{\pi} \int_0^{\infty} \frac{db}{b} \sin(b a_T)
\exp\left({-S(\bar{b}/2)}\right) \times \\ \times \int_0^1 dx_1
\int_0^1 dx_2
\left[f_q\left(x_1,(2/\bar{b})^2\right)f_{\bar{q}}\left(x_2,
    (2/\bar{b})^2 \right) +q \leftrightarrow \bar{q} \right]
\delta(M^2 - x_1 x_2 s)\,.
\end{multline}
The function $S(\bar{b}/2)$ is just the radiator $R(\bar{b})$ without
the $N$ dependent anomalous dimension terms which have been absorbed
into the pdfs:
\begin{equation}
S(\bar{b}/2) = 2 C_F \int_{(2/\bar b)^2}^{M^2}
\frac{dk_t^2}{k_t^2}\frac{\alpha_s(k_t^2)}{2 \pi} 
\left (\ln \frac{M^2}{k_t^2} -\frac{3}{2}\right )\,.  
\end{equation}
As we just pointed out this function coincides with the corresponding
function in the $p_T$ case.  All differences between $a_T$ and $p_T$
thus arise from the fact that one has to convolute the resummed
exponent and pdfs with a sine function representing the constraint on
a single component of $k_t$, rather than a Bessel function
representing a constraint on both components of the $k_t$.  Having
achieved the resummation we shall next expand our resummed result to
order $\alpha_s^2$ and compare its expansion to fixed-order results to
non-trivially test the resummation.

\section{Comparison to fixed-order results}
In order to non-trivially test the resummation we have the option of
expanding the results to order $\alpha_s^2$ (i.e.~up to two-loop
corrections to the Born level) and testing the logarithmic structure
against that emerging from fixed-order calculations. Since the results
for the $p_T$ distribution are already well-known and since many terms
are common to the $a_T$ and $p_T$ resummed results it is most
economical to provide a prediction for the difference between the
$a_T$ and $p_T$ variables. To be precise we already identified a
leading-order result for the difference between cross-sections
involving $a_T$ and $p_T/2$ in eq.~\eqref{eq:lodiff}. In this section
we shall derive this difference at NLO level and compare to
fixed-order estimates.

Let us consider the resummed results for the $a_T$ and $p_T/2$ cases.
We remind the reader of the well-known result for the $p_T$ variable
by expressing the integrated cross-section for events with $p_T/2$
below a fixed value $\epsilon M$:
\begin{equation}
 \label{eq:sigma-pt-def}
 \Sigma\left(p_T/2 \right)|_{p_T/2=\epsilon M} = \left( 1+C_1(N) \,
   \frac{\alpha_s}{2\pi} \right) 
 \tilde{\Sigma}^{(0)}(N) 
 \int_0^{\infty} \!\!db\, 2M \epsilon \> J_1(b \,2 M \epsilon)\,
 e^{-R\left(\frac{\bar{b}}{2}\right)}\>,
\end{equation}
The above result is expressed in moment space and we have additionally
provided a multiplicative coefficient function $\left( 1+C_1(N)\,
  \alpha_s/(2\pi)\right)$, so that the result accounts also for
constant terms at leading order. This form of the resummation is
correct up to NNLL accuracy in the cross-section whereas the pure
resummed result without the multiplicative constant piece is correct
to NLL accuracy in the resummed exponent~\cite{CTTW}.  Thus with the
constant $C_1$ in place the resummation should guarantee at order
$\alpha_s^2$ terms varying as $\alpha_s^2 L^4$, $\alpha_s^2 L^3$ as
well as the $\alpha_s^2 L^2$ term which partially originates from a
``cross-talk'' between the $\alpha_s C_1$ term and the $\alpha_s L^2$
term in the expansion of the exponent.\footnote{This form of the
  result we use is an oversimplification since we consider only the
  piece of the $\mathcal{O}(\alpha_s)$ constant which is associated to
  the annihilation channel. In principle one should also include the
  constant arising from the Compton channel but this is identical to
  the corresponding constant for the $p_T$
  variable~\cite{GrazziniDeFlorian} and it is straightforward to show
  that its effects cancel to the accuracy we need for the result we
  derive below for the difference between $a_T$ and $p_T/2$
  variables.}

The equivalent result for the $a_T$ variable is
\begin{equation}
 \label{eq:sigma-at-def}
 \Sigma\left(a_T\right)|_{a_T=\epsilon M} = \left( 1+\bar{C}_1(N)\, 
   \frac{\alpha_s}{2\pi} \right) 
 \tilde{\Sigma}^{(0)}(N) \frac{2}{\pi}\int_0^{\infty} \!\frac{db}{b}\,  
 \> \sin\left(b M \epsilon \right) e^{-R\left(\frac{\bar{b}}{2}\right)}\>,
\end{equation}
where $\bar{C}_1$ is the constant for the $a_T$ variable. We shall
first expand the resummation to order $\alpha_s$ and consider the
difference between $a_T$ and $p_T/2$. First we express the radiator in
the standard notation~\cite{CTTW}
\begin{equation}
-R\left (\bar{b}/2 \right ) = \sum_{n=0}^{\infty} \sum_{m=0}^{n+1}G_{nm}\, \bar{\alpha_s}^n L^m\,,\qquad
L=\ln \left (\bar{b}^2
  M^2 /4 \right)\,,
\end{equation} 
with $\bar{\alpha_s} = \alpha_s/2\pi$. Having done so we expand the
resummed exponent so that to order $\alpha_s$ we can write for the
$p_T$ variable

\begin{multline}
  \Sigma\left(p_T/2 \right)|_{p_T/2=\epsilon M} = \left( 1+C_1(N)
    \,\bar{\alpha_s}
  \right) \tilde{\Sigma}^{(0)}(N) \int_0^{\infty} \!\!db\, 2 M \epsilon \, J_1(b \,2 M \epsilon) \,\times \\
  \times \left(1+G_{11} \bar{\alpha_s}L+G_{12}\bar{\alpha_s}L^2 +
    \mathcal{O}(\alpha_s^2)\right),
\end{multline}
where we replaced the resummed exponent by its expansion to order
$\alpha_s$.

Carrying out the $b$ integral  yields
\begin{equation}
 \Sigma\left(p_T/2 \right)|_{p_T/2=\epsilon M} = \left( 1+C_1(N) \bar{\alpha_s} 
\right) \tilde{\Sigma}^{(0)}(N) \left(1+G_{11} \bar{\alpha_s} \ln 
\left [\frac{1}{4 \epsilon^2} \right ]+G_{12} \bar{\alpha_s} \ln^2 \left [\frac{1}{4 \epsilon^2} \right ] \right),
\end{equation}
where for the moment we do not insert the explicit forms of the
$G_{nm}$ coefficients.

Repeating the exercise for the $a_T$ variable one obtains
\begin{multline}
  \Sigma (a_T)|_{a_T=\epsilon M}= \left( 1+\bar{C}_1(N) \bar{\alpha_s}
  \right) \tilde{\Sigma}^{(0)}(N) \times \\
  \times\left(1+G_{11} \bar{\alpha_s} \ln \left [\frac{1}{4
        \epsilon^2} \right ]+G_{12} \bar{\alpha_s} \ln^2 \left
      [\frac{1}{4 \epsilon^2} \right ] +G_{12}\, \bar{\alpha_s}\,
    \frac{\pi^2}{3}\right)\,,
\label{eq:atexpan}
\end{multline}
where we labelled the constant piece as $\bar{C}_1$ to distinguish it
from that for the $p_T$ variable. Constructing the difference at
$\mathcal{O}(\alpha_s)$ between the $a_T$ and $p_T/2$ variables we
find that all the logarithms cancel and we obtain
\begin{equation}
  \Sigma (a_T)|_{a_T=\epsilon M}-\Sigma\left (p_T/2 \right)|_{p_T/2=\epsilon M}=\tilde{\Sigma}^{(0)}(N)\,\bar{\alpha_s}\, \left(\bar{C}_1(N)-C_1(N)+G_{12}\, \frac{\pi^2}{3} \right).
\label{eq:lodiff2}
\end{equation}

The value of the resummation coefficient $G_{12}$ can be obtained from
eq.~\eqref{eq:g2} by expanding the result in powers of $\lambda$ from
which we find $G_{12}=-C_F$.  Comparing this result with our explicit
leading order calculation eq.~\eqref{eq:lodiff} we find that $C_1 =
\bar{C}_1$. 

Next we carry out the expansion of our resummation to order
$\alpha_s^2$ and construct the difference from $p_T/2$ at this
order. We shall then compare our expectation with MCFM. Expanding the
radiator to order $\alpha_s^2$ one gets
\begin{equation}
  \begin{split}
    e^{-R\left(\frac{\bar{b}}{2}\right)} &=
    1+\bar{\alpha_s}\left(G_{11}L+G_{12}L^2\right)\\
    &+
    \bar{\alpha_s^2}\left(\frac{G_{11}^2 L^2}{2}+G_{22}
      L^2+G_{11}G_{12}L^3+G_{23}L^3 +\frac{G_{12}^2 L^4}{2} \right).
  \end{split}
\end{equation}

Retaining only the $\mathcal{O}(\alpha_s^2)$ terms we can once again
carry out the $b$ space integrals as before and in particular the new
integrals that appear at this order are
\begin{align}
  I_{a_T}^{(p)} &= \frac{2}{\pi} \int_0^\infty \frac{db}{b} \sin (b M \epsilon) \ln^{p} \left(\bar{b}^2 \frac{M^2}{4} \right)\,, \nonumber \\
  I_{p_T}^{(p)} &= \int_0^{\infty} \!db \, 2 M \epsilon\, J_1 (b \,2 M \epsilon) \ln^{p}
  \left(\bar{b}^2 \frac{M^2}{4} \right)\,,
\end{align}
with $p=3,4$. Carrying out the above integrals with $p=3$ is
straightforward and the difference between the integrals for the $p_T$
and $a_T$ case with $p=3$ produces only an $\alpha_s^2 \ln
1/\epsilon^2$ term apart from constant pieces. Such terms are beyond
the accuracy of our resummation which ought to guarantee only terms as
singular as $\alpha_s^2 \ln^2 1/\epsilon^2$ in the cross-section and
hence to our accuracy there will be no contribution for $p=3$ for the
difference between $a_T$ and $p_T/2$.

\FIGURE{
  \epsfig{file=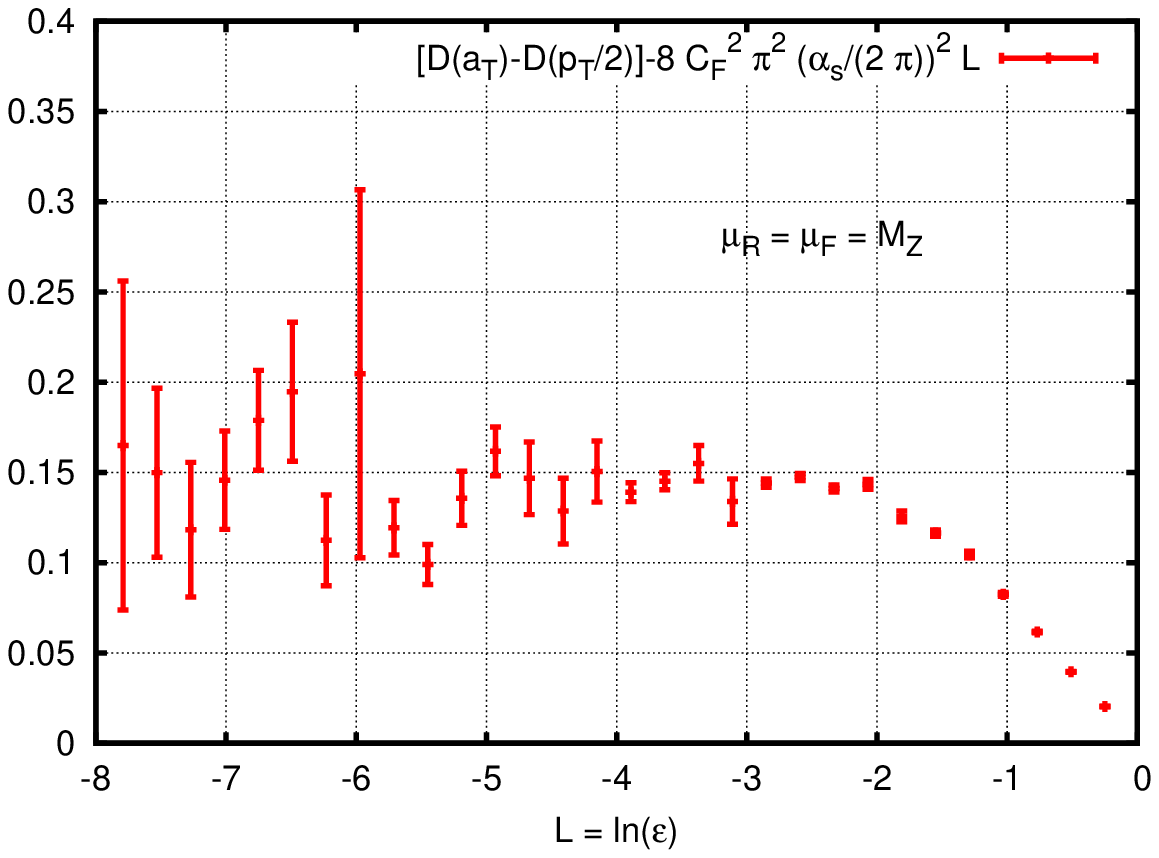,width=.9\textwidth}
  \caption[Example of figure]{Made with \tt\ttbs FIGURE.}%
  \caption{The difference between $D(a_T)$ and $D(p_T/2)$, defined in
    eq.~(\ref{eq:twoloop-diff}) with the subtraction of the computed
    logarithmic enhanced term in eq.~(\ref{eq:twoloop}).  }
  \label{fig:sigmadif-nlo-fine}
}

The situation changes when we consider the $p=4$ integrals. All
relevant logarithms cancel between $a_T$ and $p_T/2$ once again except
a term varying as $\alpha_s^2 \ln^2 \frac{1}{\epsilon^2}$. To be
precise considering only the order $\alpha_s^2$ terms one obtains
\begin{equation}
\label{eq:twoloop}
\begin{split}
  \Sigma(a_T)|_{a_T=\epsilon M}-\Sigma\left (p_T/2 \right)|_{p_T/2=\epsilon M} &= 
  \frac{G_{12}^2}{2}\, \times 2 \pi^2 \bar{\alpha_s^2}\, \ln^2 \frac{1}{\epsilon^2}\times \tilde\Sigma^{(0)}(N)+\mathcal{O}(\alpha_s^2 L)\\
&= \,\pi^2\, \bar{\alpha_s}^2\, C_F^2\,\ln^2 \left(\frac{1}{\epsilon^2} \right) \times \tilde\Sigma^{(0)}(N).
\end{split}
\end{equation}
To convert the result above back into $\tau$ space from Mellin space
is straightforward as one just inverts the Mellin transform for
$\tilde{\Sigma}^{(0)}$ to yield the Born-level quantity $\Sigma^{(0)}$
as the multiplicative factor.

Once again the above result can be tested against the results from
MCFM.  We consider the difference in the differential distributions
(derivative with respect to $\ln \epsilon$ of the appropriate
integrated cross-sections) for $a_T$ and $p_T/2$ as a function of $\ln
\epsilon$
\begin{equation}
\label{eq:twoloop-diff}
  D(a_T)|_{a_T=\epsilon M }-D\left (p_T/2 \right)|_{p_T/2=\epsilon M}=
  \frac{1}{\Sigma^{(0)}}\,\left[\frac{d\Sigma(a_T)}{d\ln \epsilon}|_{a_T=\epsilon M }-\frac{d\Sigma\left (p_T/2 \right)}{d\ln\epsilon}|_{p_T/2=\epsilon M}\right]\,.
\end{equation}
Our prediction for this difference can be obtained by taking the
derivative with respect to $\ln \epsilon$ of the RHS of
eq.~\eqref{eq:twoloop}. Subtracting this prediction from the MCFM
results should yield at most constant terms arising from the
logarithmic derivative of formally subleading $\alpha_s^2 \ln\epsilon$
terms.  That this is the case can be seen from
fig.~\ref{fig:sigmadif-nlo-fine} where we note that at sufficiently
small values of $\epsilon$ the difference between MCFM and our
prediction tends to a constant.

\section{Discussion and conclusions}
Before concluding we should comment on the resummed result
eq.~\eqref{eq:resummed}. First we note that there is the usual issue
that is involved with $b$ space resummation of the large and small $b$
behaviour of the integrand in that the resummed exponent diverges in
both limits. The small $b$ region is conjugate to the large $k_t$
regime which is beyond the jurisdiction of our resummation. At
sufficiently large $b$ on the other hand we run into non-perturbative
effects to do with the Landau pole in the running coupling. These
issues can be resolved by modifying the radiator such that the
perturbative resummation is not impacted. For instance the strategy
adopted in ref.~\cite{BozziHiggs} was to replace the resummation
variable $b$ by another variable $b^*$ which coincides with $b$ in the
large $b$ limit but at small $b$ ensures that the radiator goes
smoothly to zero. Likewise to regulate the Landau pole a cut-off was
placed in the large $b$ region of integration in the vicinity of the
Landau pole and it was checked that varying the position of the
cut-off had no impact on the resummation. Other prescriptions can be
found for instance in~\cite{DokMarWeb}.

As far as the behaviour of the resummed cross-section and consequently
the corresponding differential distribution is concerned the
difference from the $p_T$ distribution is solely due to the
convolution of the resummed $b$ space function with the $\sin(b)$
function rather than a Bessel function. As was explained in detail in
ref.~\cite{BMS} the result of convolution with a sine function
produces a distribution that does not have a Sudakov peak. The
physical reason for this is that a small value of $a_T$ can be
obtained by two competing mechanisms. One mechanism is Sudakov
suppression of gluon radiation and this is encapsulated to NLL
accuracy by the resummed exponent. The other mechanism is the
vectorial cancellation of contributions from arbitrarily hard
emissions which in this case involves cancellation only of a single
component of $k_t$ transverse to the lepton axis. This mechanism is
represented by the presence of the sine function while a
two-dimensional constraint such as that for the $p_T$ variable is
represented by a Bessel function. In the case of one dimensional
cancellation such as for the $a_T$ as well as for instance for the
dijet $\Delta \phi$ variable~\cite{BanDasDel} the cancellation
mechanism dominates the Sudakov suppression mechanism before the
formation of the Sudakov peak while for the $p_T$ case the vectorial
cancellation sets in as the dominant mechanism after the formation of
the Sudakov peak. Thus for the $a_T$ distribution one sees no Sudakov
peak but the distribution rises monotonically to a constant value as
predicted by eq.~\eqref{eq:resummed}.

To conclude, in this paper we have carried out a theoretical study
based of a variable, $a_T$ proposed in ref.~\cite{wyattvester} as an
accurate probe of the low $p_T$ region of the $Z$ boson $p_T$
distribution. Having accurate data on the $a_T$ well into the low
$a_T$ domain will be invaluable in pinning down models of the
non-perturbative intrinsic $k_t$ and may lead to firmer conclusions on
aspects such as small-$x$ broadening of $p_T$
distributions~\cite{BLNY} than have been reached at present with the
$p_T$ variable. In this respect it may also be of interest to examine
theoretically the power corrections to the $a_T$ distribution along
the same lines as for the $p_T$ case~\cite{Smye} and hence to examine
theoretically whether the $a_T$ and $p_T$ ought to have identical
non-perturbative behaviour.  This is once again work in progress.

Before any such conclusions can be arrived at however, it is of vital
importance to have as accurate a perturbative prediction as possible
to avoid misattributing missing perturbative effects to other
sources. The most accurate perturbative prediction one can envisage
for the $a_T$ case is one where resummation of large logarithms in
$a_T$ is supplemented by matching to fixed-order corrections up to the
two-loop level. In this paper we have carried out the first step by
resumming to NLL accuracy the $a_T$ distribution and checking our
resummation by comparing to the logarithmic structure in exact fixed
order calculations. We envisage that it should be possible to actually
extend the accuracy of the resummed calculation to the NNLL level
which has already been achieved for the $p_T$ variable~\cite{Catani}
and this is an avenue for future development. In any case our current
prediction matched to fixed-order estimates from MCFM should already
enable accurate phenomenological investigation alongside forthcoming
Tevatron data~\cite{wyattvester}.  As part of an article in progress
we plan to carry out the matching and a detailed phenomenological
study of the $a_T$ distribution which we anticipate will shed more
light on issues relevant to physics at the LHC in the near future.

\section*{Acknowledgements}
We wish to thank the authors of ref.~\cite{wyattvester} for informing
us about their experimental study of the $a_T$ distribution. One of us
(R.D.) would like to thank the Universit\`a degli Studi di
Milano-Bicocca and INFN, Sezione di Milano-Bicocca for generous
financial support and kind hospitality during the course of this work.

\appendix
\section{Born level result}
\label{sec:Born}
Here we explicitly compute the leading-order and the real part of the
next-to-leading order contribution to the $a_T$ integrated cross
section defined in eq.~(\ref{eq:intcross}).

At leading order $\Sigma^{(0)}(a_T)$ is just the Born cross section:
\begin{multline}
  \Sigma^{(0)}(a_T) = \int_0^1 dx_1 \int_0^1 dx_2\,\left [ f_q(x_1)
    f_{\bar{q}}(x_2) + q \leftrightarrow \bar{q} \right ] \times \\
  \times \int d\Phi(l_1,l_2)\> \mathcal{M}^2_{\mathrm{DY}}(l_1,l_2)\,
  \delta\left (M^2-2\,l_1.l_2 \right) \>,
\end{multline}
where $\mathcal{M}^2_{\mathrm{DY}}$ is the Born matrix element
reported in eq.~(\ref{eq:M2DY}). To this end we look at the
Lorentz-invariant phase-space which can be written as
\begin{equation}
\label{eq:lips}
\int d\Phi(l_1,l_2) = \frac{1}{2 \hat{s}}\int \frac{d^3l_1}{2(2 \pi)^3 l_{10}} 
\frac{d^3 l_2}{2(2 \pi)^3 l_{20}} 
(2 \pi)^4 \delta^4 (p_1+p_2-l_1-l_2)\,,
\end{equation} 
where we included in addition to the usual two-body phase space a
delta function corresponding to holding the invariant mass of the
lepton-pair at $M^2$ and $\hat{s}$ is the partonic centre of mass
energy squared $\hat{s} = s \,x_1 x_2$.

We parameterise the four vectors of the incoming partons and outgoing
leptons as below (in the lab frame)
\begin{align}
\label{eq:vec}
p_1 &= \frac{\sqrt{s}}{2} x_1 \,(1,0,0,1) , \\ \nonumber
p_2 &= \frac{\sqrt{s}}{2} x_2 \,(1,0,0,-1), \\ \nonumber
l_1 &= l_T \left(\cosh y,1,0,\sinh y \right),
\end{align}
with $l_2$ being fixed by the momentum conserving delta function.

In the above $\sqrt{s}$ denotes the centre of mass energy of the
incoming hadrons while $x_1$ and $x_2$ are momentum fractions carried
by partons $p_1$ and $p_2$ of the parent hadron momenta, while $l_T$
and $y$ are the transverse momentum and rapidity of the lepton with
respect to the beam axis and we work in the limit of vanishing lepton
masses.  In these terms we can express eq.~(\ref{eq:lips}) as (after
integrating over $l_2$ using the momentum conserving delta function)
\begin{equation}
 \int d\Phi(l_1,l_2)= \frac{1}{2\hat{s}}\int \frac{l_T dl_T dy}{4 \pi} 
 \delta \left((p_1+p_2-l_1)^2\right)\,,
\end{equation}
where we have carried out an irrelevant integration over lepton
azimuth. Note that the factor $\delta \left((p_1+p_2-l_1)^2\right)$
arises from the vanishing invariant mass of lepton $l_2$.

In order to obtain the full Born result we need to fold the above
phase space with the parton distribution functions\footnote{In order
  to avoid excessive notation we do not explicitly indicate the sum
  over incoming parton flavours which should be understood.} and the
squared matrix element for the Drell-Yan process to finally obtain
\begin{multline}
  \label{eq:sig0}
  \Sigma^{(0)}(M^2) = \int_0^1 dx_1 \,f(x_1) \int_0^1 dx_2  \,f(x_2) \,\times \\
  \times \frac{1}{2\hat{s}}\int \frac{l_T \,dl_T \,dy}{4 \pi} \delta
  \left (s \, x_1 x_2+t_1+t_2 \right) \delta \left
    (M^2-x_1x_2s \right) \, \mathcal{M}^2_{\mathrm{DY}}.
\end{multline}
where we used $\left (p_1+p_2-l_1 \right)^2 = s x_1 x_2 +t_1+t_2$.

We next evaluate the squared matrix element
$\mathcal{M}^2_{\mathrm{DY}}$ in eq.~(\ref{eq:M2DY}) in terms of the
phase space integration variables:
\begin{equation}
\label{eq:t1t2}
t_1 = - 2 p_1.l_1  =-\sqrt{s} \,x_1 \,l_T  \,e^{-y}\,, \qquad
t_2 = -\sqrt{s}\, x_2\, l_T\, e^{y} \>.
\end{equation}
Inserting these values of $t_1$ and $t_2$ in eq.~\eqref{eq:sig0} we
use the constraint 
\begin{equation}
\delta(s\, x_1x_2+t_1+t_2) = \delta\left (s\,x_1
  x_2 -\sqrt{s}\,l_T \left(x_2 \,e^{y}+x_1\, e^{-y} \right) \right)\,,  
\end{equation}
to carry out the integration over $l_T$ which gives
\begin{equation}
  \frac{1}{8 \pi s} \int_0^1 dx_1  \, f(x_1) \int_0^1 dx_2 \,f(x_2) 
  \,\delta \left(M^2-x_1 x_2 s\right) \, 
  \frac{dy}{\left (x_2 e^{y}+x_1 e^{-y} \right)^2}\, \mathcal{M}^2_{\mathrm{DY}}\,,
\end{equation}
where in evaluating $\mathcal{M}^2_{\mathrm{DY}}$ one needs to use
$l_T =\sqrt{s}\, x_1 x_2/(x_2 e^{y}+x_1e^{-y})$ , which yields using
\eqref{eq:t1t2}
\begin{align}
t_1^2 &= x_1^2 e^{-2y} \frac{M^4}{\left(x_2 e^y+x_1e^{-y}\right)^2} \,,
\\ \nonumber
t_2^2 &= x_2^2 e^{2y} \frac{M^4}{\left(x_2 e^y+x_1e^{-y}\right)^2} \,.
\end{align}

Using the above to evaluate $\mathcal{M}^2_{\mathrm{DY}}$ in
\eqref{eq:M2DY} we obtain
\begin{equation}
  \Sigma^{(0)}(M^2) = \frac{\mathcal{G}}{N_c}
  \frac{M^4}{\pi s} \int_0^1 dx_1 \,f(x_1) 
  \int_0^1 dx_2\,f(x_2) \,\delta \left(M^2-x_1 x_2 s\right) 
  \int dy \,{\mathcal{F}}(x_1,x_2,y),
\end{equation} 
where we introduced 
\begin{equation}
  \mathcal{F}\left ( x_1,x_2,y \right) = 
  A_l\, A_q\, \frac{x_1^2e^{-2y}+x_2^2e^{2y}}{\left(x_2e^y+x_1e^{-y} \right)^4}+
  B_l\, B_q\,  \frac{x_1^2e^{-2y}-x_2^2e^{2y}}{\left(x_2e^y+x_1e^{-y} \right)^4}\,.
\end{equation}
Integrating the angular function $\mathcal{F}$ over rapidity over the
full rapidity range\footnote{We can straightforwardly adapt the
  calculation to include the experimental acceptance cuts when
  available.} one finds as expected that the parity violating
component proportional to $B_l\, B_q$ vanishes and the result is
$A_l\, A_q/(3 x_1 x_2)$.  Thus the final result is (using $x_1 x_2
=M^2/s$)
\begin{equation}
  \Sigma^{(0)}(M^2) = \mathcal{G}\frac{M^2}{3 \pi}\, \frac{A_l \,A_q}{N_c} 
  \int_0^1 dx_1\,f(x_1) \int_0^1 dx_2 \,f(x_2) \>
  \delta \left (M^2-x_1 x_2s\right).
\end{equation}

We now compute the LO QCD result by considering the emission of a
gluon in the Drell-Yan (QCD annihilation) process as well as the
contribution of the quark-gluon (QCD Compton) scattering process. Thus
we consider the reaction $p_1+p_2 = l_1+l_2+k$ where $k$ is the
emitted gluon in the Drell-Yan process and a quark/anti-quark for the
Compton process.

The squared matrix elements at this order for the annihilation and
Compton processes are reported in appendix~\ref{sec:app-lo} and we
shall use those results in what follows below. 

We explicitly parameterise the momentum $k$ as below
\begin{equation}
k = k_t \left (\cosh y_k, \cos \phi, \sin \phi, \sinh y_k \right).
\end{equation}
The parameterisation of the other particles four-momenta is as in the
Born case eq.~\eqref{eq:vec}. One now has to integrate the squared
matrix elements over a three body final state and to this end we
introduce the usual Mandelstam invariants
\begin{equation}
  \label{eq:uhat-that}
  \hat{u} = -2 p_1.k = -\sqrt{s}\, x_1\, k_t\, e^{-y_k} \,,\qquad 
  \hat{t} = -2 p_2.k = -\sqrt{s}\, x_2\, k_t\, e^{y_k}  \,.
\end{equation}
The Lorentz invariant phase space is now
\begin{equation}
\int d\Phi(l_1,l_2,k) = \frac{1}{2 \hat{s}}\int \frac{d^3l_1}{2(2 \pi)^3 l_{10}} 
\frac{d^3 l_2}{2(2 \pi)^3 l_{20}} \frac{d^3k}{2(2\pi)^3 k_0}(2 \pi)^4 
\delta^4 (p_1+p_2-l_1-l_2-k) \,.
\end{equation}

Following the same procedure as in the Born case we perform the
trivial integration over $l_2$ and obtain the leading order QCD
correction to the Born result~\eqref{eq:sig0} (for the moment we are
considering just real emission terms indicated below by the label $r$)
\begin{multline}
  \Sigma^{(1)}_r(M^2) =
  \sum_{i=A,C} \int_0^1 dx_1 \,f(x_1)\int_0^1 dx_2 \, f(x_2) \,\times \\
  \times \frac{1}{2\hat{s}}\int \frac{l_T dl_T dy}{4 \pi}
  \frac{d^3 k}{2(2\pi)^3k_0}\delta \left (M^2+t_1+t_2 +2 l_1.k\right) \,\times\\
  \times \delta \left (M^2- x_1 x_2 s \left(1-\frac{2
        p_1.k}{\hat{s}}-\frac{2p_2.k}{\hat{s}}\right)\right)
  \mathcal{M}^2_{i}\,.
\end{multline}
Here the factor $\left(1-\frac{2
    p_1.k}{\hat{s}}-\frac{2p_2.k}{\hat{s}}\right)$ accounts for the
energy-momentum carried off by the radiated parton $k$ while the index
$i=A$ pertains to the QCD annihilation process while $i=C$ indicates
the QCD Compton process.  Noting that one has as before $t_1
=-\sqrt{s}\,x_1 l_T e^{-y}, t_2 = -\sqrt{s}\, x_2 l_T e^{y}$ and
additionally $2 \,l_1.k = 2 \,l_T k_t \left(\cosh(y-y_k)-\cos\phi
\right)$, we can use the constraint $\delta \left (M^2+t_1+t_2+2l_1.k
\right)$ to integrate over $l_T$ and the value of $l_T$ (and hence
$t_1, t_2$) is thus fixed in terms of other parameters:
\begin{align} 
\label{eq:paramlo}
l_T &= \frac{M^2}{\sqrt{s} \left(x_1 e^{-y}+x_2 e^{y}-2
    \frac{k_t}{\sqrt{s}} \left(\cosh(y-y_k)-\cos \phi
    \right)\right)}\,,
\nonumber \\
t_1 &= - \frac{x_1 e^{-y}\,M^2 }{\left(x_1 e^{-y}+x_2 e^{y} -2
    \frac{k_t}{\sqrt{s}} \left(\cosh (y-y_k)-\cos
      \phi\right)\right)}\,,
\end{align}
with the expression for $t_2$ the same as that for $t_1$ except that
$x_1 \,e^{-y}$ in the numerator of the above expression for $t_1$ is
to be replaced by $x_2\, e^{y}$. After integrating away the $l_T$ one
gets
\begin{multline}
  \label{eq:sig1}
  \Sigma^{(1)}_r(M^2) = \sum_{i=A,C} \int_0^1 dx_1 \, f(x_1) \int_0^1 dx_2
  \, f(x_2) \frac{1}{8\pi s}\int dy \frac{d^3 k}{2(2\pi)^3k_0}
  \,\times\\
  \times \frac{M^2}{\hat s \left(x_1 e^{-y}+x_2e^{y}-\frac{2
        k_t}{\sqrt{s}} \left(\cosh(y-y_k)-\cos \phi
      \right)\right)^2}\, \mathcal{M}^2_i\,\,\delta \left
    (M^2- z \,x_1\,x_2 s \right) \,.
\end{multline}
where we introduced $z= 1-2(p_1.k)/\hat{s}-2(p_2.k)/\hat{s}$.

We are now ready to integrate over the parton and lepton phase space
variables.  Since we are interested in the specific cross-section in
eq.~(\ref{eq:intcross}), we need to integrate over the phase space
such that the value of the $a_T$ is below some fixed value. Further we
are interested in the small $a_T$ logarithmic terms so that we
consider the region $a_T/M \ll 1$.

To avoid having to explicitly invoke virtual corrections we shall
calculate the cross-section for all events {\emph{above}} $a_T$ and
subtract this from the total ${\mathcal{O}} \left (\alpha_s \right)$
result $\Sigma^{(1)}(M^2)$ which can be taken from the
literature~\cite{AEM}:
\begin{equation}
  \Sigma^{(1)}(a_T,M^2) = \Sigma^{(1)}(M^2) -\Sigma^{(1)}_c (a_T,M^2)\,,
\end{equation}
where we shall calculate $\Sigma^{(1)}_c (a_T,M^2) = \int_{a_T}
\frac{d\sigma}{d a_T' dM^2} d a_T'$.

Moreover since we are interested in just the soft and/or collinear
logarithmic behaviour we can use the form of the $a_T$ in the
soft/collinear limit derived in the previous section.  Thus we
evaluate the integrals in eq.~\eqref{eq:sig1} with the constraint
$\Theta \left(k_t |\sin \phi| -a_T \right)$. In order to carry out the
integration let us express the parton phase space in terms of rapidity
$y_k$, $k_t$ and $\phi$.  Thus we have
\begin{equation}
\label{eq:psg}
\begin{split}
  \int \frac{d^3k}{2(2\pi)^3k_0}
  & =\int \frac{k_t dk_t d y_k d\phi}{2(2\pi)^3}\\
  & = \left(\frac{M^2}{16 \pi^2} \right) \int\frac{d \phi}{2\pi} \int
  \!dy_k \int \frac{dz\,d \Delta}{\sqrt{(1-z)^2-4z \Delta}} 
  \,\left[\delta(y_k-y_+) + \delta(y_k-y_-)\right]\,,
\end{split}
\end{equation}
where we used 
\begin{equation}
z = 1-\frac{k_t}{\sqrt{s} x_2} e^{-y_k}-\frac{k_t}{\sqrt{s}x_1}e^{y_k}\,, 
\end{equation}
which follows from the definition of $z$ and where we also introduced
the dimensionless variable $\Delta =k_t^2/M^2$. A fixed value of $z$
corresponds to two values of the emitted gluon rapidity
\begin{equation}
  \label{eq:ypm}
  y_{\pm} = \ln\left[\frac{\sqrt s\, x_1}{2\,k_t}
    \left((1-z)\pm \sqrt{(1-z)^2-4 z\Delta}\right)
  \right]\,.
\end{equation}
The requirement that the argument of the square root in
eq.~(\ref{eq:ypm}) be positive sets an upper bound for the $z$
integration, specifically $z<1-2(\sqrt{\Delta+\Delta^2}-\Delta)$. In
the limit $\Delta\to 0$, up to corrections of order $\Delta$, this
bound reduces to $z<1-2\sqrt{\Delta}$.

Having obtained the phase space in terms of convenient variables we
need to write the squared matrix elements in terms of the same. We
first analyse the QCD annihilation correction and next the Compton
piece.  In the annihilation contribution one has singularities due to
the vanishing of the invariants $\hat{t}$ and $\hat{u}$ with the
$1/(\hat{t}\hat{u})$ piece contributing up to double logarithms due to
soft and collinear radiation by either incoming parton and the $1/\hat
t$ and $1/\hat{u}$ singularities generating single logarithms. The
double logarithms arise from low energy and large rapidity emissions
(soft and collinear emissions) while the single-logarithms from
energetic collinear emissions, hence it is the small $k_t$ limit of
the squared matrix elements that generates the relevant logarithmic
behaviour. Thus we write the squared matrix element
$\mathcal{M}^2_{A}$ in eq.~\eqref{eq:dy} in terms of the variables
$\Delta$ and $z$ and then find the leading small $\Delta$ behaviour.
Specifically in the $\Delta \to 0$ limit, considering only the $1/\hat
t$ singular piece, the factor appearing in \eqref{eq:sig1} has the
following behaviour (keeping for now only the $A_l A_q$ piece of the
matrix element):
\begin{multline}
  \label{eq:ann-approx}
  \frac{1}{\hat{s}}\, \mathcal{M}^2_A\, \frac{M^2}{s \left(x_1
      e^{-y}+x_2e^{y}-\frac{2 k_t}{\sqrt{s}} \left(\cosh(y-y_k)-\cos
        \phi \right)\right)^2}
  \approx \\
  16\,g^2\,\mathcal{G}\,\frac{A_l\, A_q}{N_c}\, \frac{C_F}{\Delta}\,
  (1+z^2)\,\frac{M^2}{s}\,\frac{x_1^2 \,e^{-2y}+x_2^2 \,z^2
    \,e^{2y}}{\left (x_1 \,e^{-y}+ x_2 \,z \,e^{y}\right)^4}\,.
\end{multline}
Performing the integral over all rapidities $y$ of the lepton, the
above factor produces $16 g^2 \mathcal{G} \frac{A_l A_q}{N_c}
\frac{C_F}{\Delta} \frac{1+z^2}{3}$.  The corresponding $B_l B_q$
piece of the squared matrix element vanishes upon integration over all
rapidities.

Since the $1/\hat u$ singular term, after integration over all lepton
rapidities, gives us the same result as that arising from
eq.~(\ref{eq:ann-approx}), we can write for the annihilation process
(using eqs.~\eqref{eq:sig1}, \eqref{eq:psg}) and $g^2 = 4 \pi
\alpha_s$
\begin{multline}
  \Sigma^{(1)}_{A,c} = \mathcal{G} \frac{A_lA_q}{N_c} \int dx_1 dx_2
  f(x_1)f(x_2) \frac{M^2}{3 \pi} \int_0^1\frac{d\Delta}{\Delta}
  \int_0^{1-2\sqrt{\Delta}} \!\!\!\!dz\,\,\delta \left(M^2-x_1 x_2 z\,
    s\right) \\ \int_0^{2\pi} \frac{d\phi}{2\pi} \, C_F \frac{\alpha_s}{2\pi}
  \frac{2\left(1+z^2\right)}{\sqrt{(1-z)^2-4z \Delta}} \Theta
  \left(\sqrt\Delta |\sin \phi| -\frac{a_T}{M} \right)\,.
\end{multline}

Following the same procedure for the Compton process one finds instead
\begin{multline}
  \frac{1}{\hat{s}} \mathcal{M}^2_C \frac{M^2}{s \left(x_1 e^{-y}+x_2e^{y}-\frac{2 k_t}{\sqrt{s}} \left(\cosh(y-y_k)-\cos \phi \right)\right)^2} \\
  \approx 16 \,g^2 \,\mathcal{G}\, \frac{A_l A_q}{N_c} \,
  \frac{T_R}{\Delta}\,(1-z) \,[z^2+(1-z)^2]
  \,\frac{M^2}{s}\,\frac{x_1^2 \,e^{-2y}+x_2^2 \,z^2 \,e^{2y}}{\left
      (x_1 \,e^{-y}+ x_2 \,z \,e^{y}\right)^4}\,,
\end{multline}
which after integration over the rapidity $y$ reduces to $16 g^2
\mathcal{G} \frac{A_l A_q}{N_c} \frac{T_R}{\Delta} (1-z)
\frac{z^2+(1-z)^2}{3}$. Thus we have for this piece
\begin{multline}
  \Sigma^{(1)}_{C,c} = \mathcal{G}\frac{A_lA_q}{N_c} \int dx_1 dx_2
  f(x_1)f(x_2) \frac{M^2}{3 \pi} \int_0^1 \frac{d\Delta}{\Delta} 
  \int_0^{1-2\sqrt{\Delta}} \!\!\!\!dz\,\,\delta \left(M^2-x_1 x_2
    z\, s\right) \\ \int_0^{2\pi} \frac{d\phi}{2\pi} \, T_R
  \frac{\alpha_s}{2\pi} \frac{(1+z)\,[z^2+(1-z^2)]}{\sqrt{(1-z)^2-4z
      \Delta}} \Theta \left(\sqrt\Delta |\sin
    \phi| -\frac{a_T}{M} \right)\,.
\end{multline}

\section{Leading Order Cross Section in the small $\mathbf{\Delta}$
  limit}
\label{sec:app-lo}
The matrix element squared for the QCD annihilation process from
ref.~\cite{AuLi} is (in four dimensions)
\begin{multline}
  \label{eq:dy}
  \mathcal{M}^2_A(l_1,l_2,k) = -16\, g^2 \,{\cal G}\, \frac{C_F}{N_c}\, M^2 
  \times \\
  \times \left\{ A_l A_q\left
      [\left(1+\frac{\hat{s}-2t_1-M^2}{\hat{t}} -
        \frac{t_1^2+t_2^2+\hat{s}\left(t_1+t_2+M^2 \right)}
        {\hat{t}\hat{u}}\right) + \left(\hat{u} \leftrightarrow
        \hat{t}, t_1 \leftrightarrow t_2 \right)\right]\right. \\
  \left.+B_l B_q \left[\left(\frac{\left(\hat{s}+2t_2+M^2\right)}
        {\hat{t}}+\frac{M^2\left(t_1-t_2 \right)}{\hat{t}\hat{u}}
      \right) -\left(\hat{u} \leftrightarrow \hat{t}, t_1
        \leftrightarrow t_2 \right)\right]\right\}\,,
\end{multline}
while for the QCD Compton process, if $p_2$ represents an incoming
gluon, one has
\begin{multline}
\label{eq:comp}
\mathcal{M}^2_C(l_1,l_2,k) = -16 \,g^2\, {\cal G} \,\frac{T_R}{N_c} M^2 
\times \\
\times \left\{ A_lA_q \left [\frac{\hat{t}-2\left
          (t_1+M^2\right)}{\hat{s}} +
      \frac{\hat{s}+2(t_1+t_2)}{\hat{t}}+\frac{2}{\hat{s}\hat{t}}
      \left(\left (t_1+t_2+M^2 \right)^2+t_1^2-t_2 M^2 \right)\right]\right. \\
  \left.+B_l B_q\left
    [\frac{2\left(t_1+M^2\right)-\hat{t}}{\hat{s}}+\frac{\hat{s}+2\left
        (t_1+t_2 \right)}{\hat{t}}-\frac{2 M^2\left(2t_1+t_2+M^2\right
      )}{\hat{s}\hat{t}} \right]\right\}\,,
\end{multline}
where we corrected small errors (after an independent recomputation of
the above) of an apparent typographical nature in the $B_l B_q$ piece
of the annihilation result.

Here we report also the electroweak coefficient constants
$\mathcal{G}$, $A_l$, $A_q$, $B_l$, $B_q$ for the case of $Z$ boson
exchange:
\begin{equation}
  \label{eq:GAB}
  \begin{split}
    &\mathcal{G}(\alpha,\theta_W,M^2,M_Z^2) = \frac{4\pi^2 \alpha^2}{\sin^4 \theta_W \cos^4 \theta_W}\frac{1}{(M^2-M_Z^2)^2+(\Gamma_Z M_Z)^2}\,,\\
    &A_f = a_f^2+b_f^2\,,\qquad B_f = 2 a_f b_f\,,\qquad (f=l,q)\,.
  \end{split}
\end{equation}
All these quantities have been taken from ref.~\cite{AuLi}, where the
reader can find analogous expressions for the case in which a virtual
photon is exchanged as well. Following the conventions of
ref.~\cite{AuLi}, we also have
\begin{equation}
  \label{eq:ab}
  \begin{split}
    a_l = -\frac{1}{4}+\sin^2\theta_W\,,&\qquad b_l = \frac 14\,,\\
    a_{u,c} = \frac{1}{4}-\frac 23\sin^2\theta_W\,,
    &\qquad b_{u,c} = -\frac 14\,,\\
    a_{d,s,b} = -\frac{1}{4}+\frac 13\sin^2\theta_W\,,
    &\qquad b_{d,s,b} = \frac 14\,.
  \end{split}
\end{equation}
The kinematical variables $\hat u$ and $\hat t$ are defined in
eq.~(\ref{eq:uhat-that}).  In the limit of small $\Delta$ both matrix
elements become collinear singular.  In the annihilation subprocess
this occurs when the emitted gluon $k$ is collinear to either $p_1$
(corresponding to $\hat u \to 0$) or $p_2$ ($\hat t \to 0$). The
singularity for $\hat u \to 0$ occurs at positive gluon rapidity
$y_k$, correspondingly the one for $\hat t \to 0$ occurs at negative
$y_k$. The matrix element for the Compton process shows only a
collinear divergence when an outgoing quark is collinear to the
incoming gluon, corresponding to $\hat t \to 0$. In the following we
compute the approximated expression of $\mathcal{M}^2_A$ and
$\mathcal{M}^2_C$ in the collinear limit $\hat t \to 0$. The remaining
collinear limit $\hat u\to 0$ of $\mathcal{M}^2_A$ gives an identical
result after integration over the lepton rapidity.

Neglecting terms of relative order $k_t$ one has
\begin{equation}
  \label{eq:smallDelta-variables}
  \begin{split}
    &l_T \simeq \frac{M^2}{\sqrt s \left(x_1 e^{-y}+ z x_2
        e^{y}\right)}\,, \qquad \hat t \simeq
    -\frac{k_t^2}{1-z}\,,\qquad
    \hat u \simeq -\frac{1-z}{z} M^2\,,\\
    & t_1 \simeq -\frac{x_1 e^{-y} M^2}{\left(x_1 e^{-y}+ z x_2
        e^{y}\right)}\,, \quad t_2 \simeq -\frac{x_2 e^{y}
      M^2}{\left(x_1 e^{-y}+ z x_2 e^{y}\right)}\,,\quad M^2 \simeq
    -(t_1 + z \,t_2)\,.
  \end{split}
\end{equation}
Substituting these expressions in eq.~(\ref{eq:dy}) and
eq.~(\ref{eq:comp}) one obtains 
\begin{equation}
  \label{eq:dy-smallDelta}
  \mathcal{M}^2_A(l_1,l_2,k) \simeq \frac{16\,g^2}{z \,k_t^2}\,{\cal G} 
  \,\frac{C_F}{N_c}
  \,(1+z^2)\, 
  \left[A_l A_q (t_1^2 + z^2 t_2^2) + B_l B_q (t_1^2 - z^2 t_2^2)\right]\,,
\end{equation}
and
\begin{equation}
  \label{eq:comp-smallDelta}
  \mathcal{M}^2_C(l_1,l_2,k) \simeq \frac{16\,g^2}{z \,k_t^2}\,{\cal G} \,\frac{T_R}{N_c}
  \,(1-z)\,[z^2+(1-z)^2] 
  \left[A_l A_q (t_1^2 + z^2 t_2^2) + B_l B_q (t_1^2 - z^2 t_2^2)\right]\,.
\end{equation}
where the collinear singularity $1/k_t^2$ has been isolated. Note that
the result is proportional to the Born matrix element in
eq.~(\ref{eq:M2DY}) with $x_2$ replaced by $z \, x_2$, indicating that
the momentum fraction of the parton entering the hard scattering has
been reduced by a factor $z$ after the emission of a collinear gluon.

\nopagebreak

\end{document}